\documentclass[twocolumn,           
               showpacs,            
               nopreprintnumbers,     
               aps,                 
               prd,          	    
               letterpaper,             
              groupeaddress,      
               nofootinbib,         
               tightenlines,        
               floats,floatfix,      
               showkeys
               ]{revtex4-1}

\usepackage{graphicx}
\usepackage{dcolumn}
\usepackage{bm}
\usepackage{amsmath}
\usepackage{amsfonts,amssymb}
\usepackage{color}
\usepackage{enumerate}
\usepackage{float}
\usepackage{subfig}
\usepackage{ulem}

\begin{document}

\title{Constraints and cosmography of $\Lambda$CDM in presence of viscosity}

\author{L. Herrera-Zamorano$^1$}
\email{lherrera31@alumnos.uaq.mx}

\author{A. Hern\'andez-Almada$^{1}$}
\email{ahalmada@uaq.mx}

\author{Miguel A. Garc\'ia-Aspeitia$^{2,3}$}
\email{aspeitia@fisica.uaz.edu.mx}

\affiliation{$^1$Facultad de Ingenier\'ia, Universidad Aut\'onoma de Quer\'etaro, Centro Universitario Cerro de las Campanas, 76010, Santiago de Quer\'etaro, M\'exico}

\affiliation{$^2$Unidad Acad\'emica de F\'isica, Universidad Aut\'onoma de Zacatecas, Calzada Solidaridad esquina con Paseo a la Bufa S/N C.P. 98060, Zacatecas, M\'exico.}

\affiliation{$^3$Consejo Nacional de Ciencia y Tecnolog\'ia, Av. Insurgentes Sur 1582. Colonia Cr\'edito Constructor, Del. Benito Ju\'arez C.P. 03940, Ciudad de M\'exico, M\'exico.}

\begin{abstract}
In this work, we study two scenarios of the Universe filled by a perfect fluid following the traditional dark energy and a viscous fluid as dark matter. In this sense, we explore the simplest case for the viscosity in the Eckart formalism, a constant, and then, a polynomial function of the redshift. We constrain the phase-space of the model parameters by performing a Bayesian analysis based on Markov Chain Monte Carlo method and using the latest data of the Hubble parameter (OHD), Type Ia Supernovae (SNIa) and Strong Lensing Systems. The first two samples cover the region  $0.01<z<2.36$. Based on AIC, we find equally support of these viscous models over Lambda-Cold Dark Matter (LCDM) taking into account OHD or SNIa.
On the other hand, we reconstruct the cosmographic parameters ($q,j,s,l$) and find good agreement to LCDM within up to $3\sigma$ CL. Additionally, we find that the cosmographic parameters and the acceleration-deceleration transition are sensible to the parameters related to the viscosity coefficient, making of the viscosity an interesting physical mechanism to modified them. 
\end{abstract}

\keywords{cosmology, viscous bulk, cosmological constrictions.}
\pacs{}
\date{\today}
\maketitle


\section{Introduction}

Nowadays, one of the challenges in cosmology is the understanding and description of the accelerated expansion phase of the Universe. Several cosmological observations give support to this phenomena; firstly it was confirmed by Supernovaes of the Type Ia (SNIa) \cite{Riess:1998} then, by the acoustic peaks of cosmic microwave background radiation (CMB) \cite{Planck:2018} and supported by 
direct measurements of the Hubble parameters (OHD) \cite{Magana:2017nfs}, Baryon acoustic oscillations (BAO) \cite{bao:2017}, and strong lensing systems (SLS) \cite{Amante:2019xao}. The simplest model, called $\Lambda$-Cold Dark Matter (LCDM), describes very well such cosmological observations and proposes a cosmological constant ($\Lambda$ or CC) characterized by an equation of state (EoS) $w=-1$ to model the accelerated expansion of the Universe and dust matter ($w=0$) to simulate the dark matter evolution at the background level. These two extra components correspond about $\sim95\%$ of the total \cite{Planck:2018}, being the rest of components associated with baryons and relativistic species like photons and neutrinos. Also, in the literature these first two ingredients are known as dark energy (DE) and dark matter (DM).

Besides its successful at large scale, LCDM presents several problems at local scales, for instance, the well known {\it missing satellite problem} that refers to the discrepancy of about 10 times more dwarf galaxies obtained by the numerical simulations based on LCDM model and the observed ones in cluster of galaxies \cite{Klypin:1999uc,Moore:1999nt}. Also, the well-known {\it core-cusp problem} \cite{NFW:1996}.
Furthermore, there are open questions concerning to the origin of the cosmological constant. In this vein, it is the {\it concordance problem} that consists in a disagreement of about $120$ orders of magnitude in the CC value measured from the Quantum Field Theory point of view and the one obtained from cosmological measurements \cite{RevModPhys.61.1,Astashenok_2012,martin2012everything}. Additionally, the {\it degeneracy problem} which afflicts also the LCDM model, refers to the inability of measuring the energy-momentum of each component, instead the total one. In other words, this implies the inability to know if the dark sector is composed by one or several components\footnote{For a review of the LCDM problems, see for instance \cite{del2017small}.}.

Several models have emerged in order to propose alternatives to the LCDM paradigm, for instance, brane-world \cite{Garcia-Aspeitia:2018fvw}, Chaplygin gases \cite{Hernandez-Almada:2018osh}, Unimodular gravity \cite{Garcia-Aspeitia:2019yod,Garcia-Aspeitia:2019yni}, among others \cite{Copeland:2006wr}, have entered into the scene as a greater contenders, resolving conundrums that the LCDM cannot, moreover, scalar fields as DM \cite{Matos:2000ss,Matos:2000ng,Matos:2000ki,UrenaLopez:2000aj}, axion \cite{Peccei:2006as,Berenji:2016jji}, etc, are an important approaches to resolve the problem of DM. In this vein, fluids with viscosity are great candidates not only to aboard the DM problem, but also, the DE problem from an unifying approach. Indeed, the viscous models could affront problems like the $H_0$ tension, and the problem associated with the matter fluctuation amplitude. This kind of models also have the characteristic of predicting an earlier transition to an accelerated phase in comparison with the standard cosmological model (see for example \cite{Avelino:2013,Atreya_2018,DiValentino:2019ffd}). This is because the causative of the Universe acceleration could be related with a dynamical DE and not a cosmological constant as states the consensus model. Moreover, we cannot ignore the recent results of the Experiment to Detect the Global EoR Signature (EDGES) which detect an excess of radiation (not predicted by the standard model) in reionization epoch, specifically in $z\approx17$ \cite{Bowman:2018yin}, which can be boarded through the viscous model scenario.

In the non perfect fluid context, there are two kind of viscosity known as shear and bulk. The shear viscosity does not play an important role at late epochs of the Universe because it does not satisfy the cosmological principle as the bulk viscosity does. Then, at late times, it is of great interest to study the bulk viscosity in any of its formalisms and there are, typically, two ways to aboard it, using the Eckart theory \cite{Eckart:1940} or the Israel-Steward (IS) theory \cite{Israel1979}. Although, in the Eckart approach the propagation of the perturbations on the non perfect fluid occurs at infinite speed, it is a simpler theory than the IS formalism, allowing to study more complex forms of the viscosity. For instance, bulk viscosity coefficient has been modelled as a constant \cite{Murphy:1973, Padmanabhan:1987, Brevik2005, Normann:2017}, polynomials \cite{Xin-He:2009, Avelino:2010, Almada:2019}, and hyperbolic \cite{FOLOMEEV200875, Almada:2019} functions. Moreover, it allows easily to explore the presence of interacting terms in the viscous fluid \cite{Almada:2020}. Because the non perfect fluid should satisfy the near equilibrium condition of thermodynamics, the pressure of the fluid must be greater than the one generated by the viscosity. Then, to alleviate this condition, it is convenient to add an extra perfect fluid such as CC or any other. It is worth mention that this assumption has a price to pay in the viscous fluids because they loss the power to describe the dynamics of the Universe by unifying the DM and the DE in an unique viscous fluid.

Regarding to IS formalism, there is a small quantity of viscosity models that has been studied. For instance, some authors  \cite{Cornejo:2013, MCruz:2017, CRUZ2017159, NCruz:2018arx, NCruz:2019, cruz2019unified} consider the viscosity coefficient as $\xi \sim \rho^s$, where $\rho$ is the energy density of the fluid and some solutions have been studied for the case $s=1/2$. Nevertheless, to solve the mentioned near equilibrium condition of thermodynamics, \cite{NCruz:2018} studies the evolution of the Universe by adding a fluid as CC to a viscous fluid (considering $\xi \sim \rho^s$). 

In this work, we study a Universe filled by two fluids under the Eckart formalism, a perfect fluid as DE mimicking the dynamics of the CC, while a non-perfect fluid as DM which is added to its EoS a viscosity term. We will assume two cases for the viscosity coefficient: a constant and a polynomial function of the redshift. It is interesting to remark that our model is reduced to LCDM model (when the relativistic species are negligibly) by turning off the viscosity. In this sense, we aim mainly to compare the cosmographic parameters of our viscous models and LCDM to understand the viscosity effects in LCDM. Additionally, we analyze correlations between the cosmography and the viscosity parameters. Hence, we perform a Bayesian Markov Chain Monte Carlo (MCMC) analysis to constrain the free model parameters using the largest samples of the observational Hubble parameter distance measurements \cite{Magana:2017nfs}, Supernovae Pantheon sample \cite{Scolnic:2017caz} and Strong Lensing Systems \cite{Amante:2019xao}. 

The manuscript is structured as follow: In Sec. \ref{sec:models}, it is presented the mathematical details of the viscous dark fluid model. Section \ref{sec:Data} describes the OHD, SNIa and SLS samples, together with the joint analysis of the three previously mentioned samples, in Sec. \ref{Sec:ConsandRes} we present the constraints and results, finally in Sec. \ref{Sec:DisCon} we give some discussions and conclusions.

\section{Viscous dark fluid model}\label{sec:models}

In what follow, we summarize briefly the two components models following the mathematical formalism described in \cite{Normann:2016,Normann:2017}.  Thus, we consider a flat Universe ($k=0$) under the Friedmann–Lema\^itre–Robertson–Walker (FLRW) metric, 
\begin{equation}\label{eq:metric}
    ds^2 = g_{\mu\nu}dx^{\mu}dx^{\nu}= -dt^2 + a(t)^2(dr^2+r^2d\Omega^2)\,,
\end{equation}
where $a(t)$ is the scale factor as function of the cosmic time and $d\Omega^2=d\theta^2+\sin^2\theta d\varphi^2$ is the solid angle. In addition, the bulk viscosity term, $\Pi$, is introduced through the energy-momentum tensor as an effective pressure $\Tilde{p}$, {\it i.e.,}
\begin{equation}\label{eq:Tmunu}
    T_{\mu\nu} = \rho u_\mu u_\nu + \Tilde{p}h_{\mu\nu}\,,
\end{equation}
being $h_{\mu\nu}=g_{\mu\nu} + u_\mu u_\nu$ and $u^{\mu}=(1,0,0,0)$ is the cuadrivelocity in the co-moving coordinate system, $\Tilde{p}= p + \Pi$ with $p$  is the total barotropic pressure of the fluids presented in the Universe. In this vein, we consider a viscous dust-like matter, with EoS $w=p/\rho=0$ coupled to a perfect fluid  behaving as the cosmological constant, {\it i.e.}, $w=-1$. Then, the Friedmann equations are
\begin{eqnarray}
&&H^2 = \frac{\kappa^2}{3}(\rho_m + \rho_{de}), \label{eq:H}\\ 
&&\dot{H} + H^2 = -\frac{\kappa^2}{6}(\rho_m + \rho_{de} + 3\Tilde{p})\,, \label{eq:Hdot} \\
&&\dot{\rho}_{m} + 3H\rho_{m} = 9\xi H^2 \,,  \label{eq:mdot}\\
&&\dot{\rho}_{de} + 3H\rho_{de} = 0\,, \label{eq:dedot}
\end{eqnarray}
where  $\kappa^2 = 8\pi G$ and $H=\dot{a}/a$ is the Hubble parameter, $\rho_m$ and $\rho_{de}$ are the energy density of the viscous matter and dark energy, respectively. The effective pressure is $\Tilde{p}= -\rho_{de} - 3\xi H$ where we have inspired the form of $\Pi$ to be  proportional to the $H$ through the bulk viscosity coefficient  $\xi=\xi(t)$ as analogy to the fluid dynamics which the viscous effects are proportional to the velocity. 

Based on the procedure presented in \cite{Normann:2016,Normann:2017}, the homogeneous solution (when $\Pi=0$) of Eqs. (\ref{eq:mdot})-(\ref{eq:dedot})  with respect to the redshift $z$ is
\begin{equation}
    \rho_h(z) = \sum_i \rho_{i0}(1+z)^{-3(w_i+1)}\,,
\end{equation}
where $i=m, de$, and $\rho_{i0}$ means the energy density at current epochs of the $i$-component. Notice that this solution corresponds to an Universe filled by two perfect fluids. Additionally, it is interesting to see that it is an approximation of the LCDM model which we have despised the relativistic species component at the background level. The viscosity effects will appear as correction terms in the general solution which is built as the sum of the homogeneous solution and the particular one. In other words, we can express the general solution as
\begin{equation}\label{eq:rho2}
    \rho(z) = \rho_h(z) [1 + u(z)]\,. 
\end{equation}
It is important to remark that this form of the general solution is a simplification coming from the idea that the viscous term is the same for all the fluids, implying that the function $u(z)$ is the same for any fluid. For more details see \cite{Normann:2016}. Then, it can be possible to find an expression for $u(z)$ in function also with the bulk viscosity coefficient, through the following differential equation, as
\begin{equation}
    -(1+z)\frac{du(z)}{dz} = 9\frac{\xi(z)}{\rho_h(z)}\sqrt{ \frac{\kappa^2}{3}\rho_h(z)[1+u(z)] }\,.
\end{equation}
By defining the dimensionless viscosity coefficient $\lambda(z) = \xi(z) H_0/\rho_{cr}$, being  $\rho_{cr}=3H_0^2/\kappa^2$ the critical density, the above expression reads as
\begin{equation}
   1+u(z) = \left[ \frac{9}{2}\int_0^z \frac{\lambda(z)\, }{(1+z)\sqrt{\Omega(z)}}dz +I_0 \right]^2 \,, 
\end{equation}
where $I_0$ is an integration constant and we have defined $\Omega(z) = \rho_h(z)/\rho_{cr} = \sum_i \Omega_{i0}(1+z)^{3(1+w_i)}$, $\Omega_{i0}=\rho_{i0}/\rho_{cr}$, being $i=m,\, de$. Now, in order to solve the above integral, we need to propose a form of $\lambda(z)$.

By considering the form $9\lambda(z) = \lambda_0 + \lambda_1(1+z)^n$, \cite{Hernandez-Almada:2019}, we have
\begin{equation}
    1+u(z) = \left[ \frac{1}{2}\int_0^z \frac{\lambda_0 + \lambda_1(1+z)^n\, }{(1+z)\sqrt{\Omega(z)}}dz + I_0 \right]^2 \,,
\end{equation}
where $\lambda_0$, $\lambda_1$ ans $n$ are free parameters. Before to solve the latter expression, it is convenient to write the dimensionless Hubble parameter as function of the redshift, $z$, defined as
\begin{equation}\label{eq:defE2}
    E(z)^2\equiv\frac{H(z)^2}{H_0^2} = \frac{1}{\rho_{cr}}\sum_i\rho_i\,,
\end{equation}
where $i=m,\,de$ and $H_0=100\,h\, {\rm km\,s^{-1}\,Mpc^{-1}}$ and $h$ is the dimensionless Hubble constant measured at current epochs. Then, by integrating and using (\ref{eq:rho2}) and (\ref{eq:defE2}), we obtain,
\begin{eqnarray} \label{eq:E2pol}
    E(z) &=& \sqrt{\Omega(z)}\left[1 + \frac{\lambda_0}{3\sqrt{\Omega_{de0}}}\operatorname{sinh}^{-1}{\left (\sqrt{\frac{\Omega_{de0}}{ \Omega_{m0}(1+z)^3}} \right )} \right. \nonumber \\
    && \left. - \frac{\lambda_0}{3\sqrt{\Omega_{de0}}} \operatorname{sinh}^{-1}{\left (\sqrt{\frac{\Omega_{de0}}{\Omega_{m0}}} \right )} \right] \nonumber \\
         & & +\sqrt{\Omega(z)} \left[  \frac{\lambda_1}{2n\sqrt{\Omega_{de0}}}(1+z)^n \times \right. \nonumber \\ 
         & & \left. \,{{}_{2}F_{1}}\left( \frac{1}{2}, \frac{n}{3}, 1+\frac{n}{3}, -\frac{\Omega_{m0}(1+z)^3}{\Omega_{de0}} \right. \right) \nonumber \\
         & & \left. - \frac{\lambda_1}{2n\sqrt{\Omega_{de0}}}\,{{}_{2}F_{1}}\left( \frac{1}{2}, \frac{n}{3}, 1+\frac{n}{3}, -\frac{\Omega_{m0}}{\Omega_{de0}}  \right)  \right]\,,
\end{eqnarray}
where we have used $E(0)=\Omega(0)=1$, $\Omega(z) = \Omega_{m0}(1+z)^3+\Omega_{de0}$, and ${}_{2}F_{1}$ is the hypergeometric function. In this work, based on the results obtained on \cite{Hernandez-Almada:2019} we will set $n=-2$. 
For $\lambda_1=0$, we have the case when the bulk viscosity coefficient is constant; in this case, we obtain 
\begin{eqnarray}\label{eq:E2cte}
    E(z) &=& \sqrt{\Omega(z)}\left[1 + \frac{\lambda_0}{3\sqrt{\Omega_{de0}}}\operatorname{sinh}^{-1}{\left (\sqrt{\frac{\Omega_{de0}}{ \Omega_{m0}(1+z)^3}} \right )} \right. \nonumber \\
        & & \left.- \frac{\lambda_0}{3\sqrt{\Omega_{de0}}} \operatorname{sinh}^{-1}{\left (\sqrt{\frac{\Omega_{de0}}{\Omega_{m0}}} \right )} \right]\,,
\end{eqnarray}
where $\sinh^{-1}$ is the inverse of hyperbolic $\sin$ function. In the latter we can observe that the standard cosmology (LCDM) is recovery when $\lambda_0=0$. It is worth to notice that for these two cases, a future singularity appears for $z\rightarrow -1$, crossing to the Phantom DE region. An evidence of this behavior it is discussed later in Sec. \ref{Sec:DisCon}, which coincide with previous studies provide by \cite{Brevik:2011mm} about a Little Rip for this kind of models.

\section{Cosmological Samples} \label{sec:Data}

In order to analyse the viscous models, we use three data samples provided by direct measurements of the Hubble parameters,  Supernovae observations and strong lensing system. This section is devoted to describe them and report the constraints of the model parameters considering each sample and also by performing a joint analysis.

\subsection{Observational Hubble parameter measurements}

Currently, the direct way of measuring the accelerated expansion of the Universe is through the Hubble parameter using the differential age tools and BAO measurements. The largest sample that include these observations is compiled by \cite{Magana:2017nfs} covering a range $0.07<z<2.36$ with $51$ points. We will refer to this sample as the Observational Hubble distance (OHD). Then, to constrain the parameter phase space, ${\bf \Theta}=(h, \Omega_m, \lambda_0, \lambda_1)$  and ${\bf \Theta}=(h, \Omega_m, \lambda_0)$ for the model when  $\lambda$ is polynomial (setting $n=-2$) and constant, respectively, we build the $\chi^2$-function to be minimize as
\begin{equation}\label{eq:chi2_ohd}
\chi^2_{OHD} = \sum_{i=1}^{51} \left( \frac{H_{th}(z_i, {\bf \Theta}) - H_{obs}^i}{\sigma_{obs}^i} \right)^2 \,.
\end{equation}
In the above expression, $H_{obs}^i$ represents the observational  Hubble parameter with its uncertainty $\sigma_{obs}^i$ at the redshift $z_i$. On the other hand, $H_{th}$ represents the theoretical expression related to the Eqs.  (\ref{eq:E2pol}) and (\ref{eq:E2cte}) for the  polynomial  and constant form of the bulk viscosity coefficient ($\lambda$), respectively.

\subsection{Type Ia Supernovae data}

Apart of the OHD sample, it is useful to include the luminosity distance measurements obtained through Type Ia Supernovae (SNIa). The largest sample, collected by Pantheon \cite{Scolnic:2017caz}, covers the redshift region $0.01 < z < 2.3$ with $1048$  measurements of the bolometric apparent magnitude. In order to compare it with our models, we compute the theoretical one as
\begin{equation}
m_{th}(z) = \mathcal{M} + 5\, \log_{10}\left[ d_L(z)/10\,pc \right]\,,
\end{equation}
where, $\mathcal{M}$ is a nuisance parameter. The quantity $d_L(z)$, known as the dimensionless luminosity distance,  is given by
\begin{equation}
d_L(z) =  (1+z) \frac{c}{H_0} \int_0^z \frac{dz'}{E(z')} \,,
\end{equation}
where $c$ is the speed of light and $E(z)$ the dimensionless Hubble parameter presented in Eqs (\ref{eq:E2pol}) and (\ref{eq:E2cte}). Then, we build the $\chi^2$-function as
\begin{equation}
\chi^2_{SNIa} = (m_{th}-m_{obs}) \cdot {\rm Cov}^{-1} \cdot (m_{th}-m_{obs})^{T}\,,
\end{equation}
where ${\rm Cov}^{-1} $ refers to the inverse of the covariance matrix and $m_{obs}$  to the observed quantity of $m$.

\subsection{Strong Lensing System}

Finally, we also use the latest compilation of the strong lensing systems (SLS) provided by \cite{Amante:2019xao} which consider only systems where the lens are early type galaxies. With a total of $205$ points, the sample covers  the redshift region  $0.0625<z_l<0.9280$ for the lens galaxy and $0.196<z_s<3.595$ for the source.  To constrain cosmological parameters, it is useful to built the chi-square  function given by \cite{Grillo:2008}
\begin{equation}
\chi^2_{SLS} = \sum_{i=1}^{205} \left( \frac{D_{th}(z_l, z_s, {\bf \Theta}) - D_{obs}^i}{\delta D_{obs}^i} \right)^2 \,.
\end{equation}
In the latter, $D_{th}$ is the theoretical angular diameter distance ratio defined by
\begin{equation}
D_{th} = \frac{D_{ls}}{D_s}\,,
\end{equation}
being the angular diameter distance to the source,
\begin{equation}
D_s = \frac{1}{1+z}\frac{c}{H_0} \int_0^{z_s} \frac{dz}{E(z)}\,.
\end{equation}
Similarly, $D_{ls}$ means the angular diameter distance between the source and the lens galaxy, i.e., the same previous equation but now evaluated in the region $z_s<z<z_l$. 
The observational counterpart,  $D_{obs}$, is built as 
\begin{equation}
D_{obs} = \frac{c^2 \theta_E}{4\pi \sigma^2}\,,
\end{equation}
where $\theta_E$ is known as the Einstein radius and $\sigma$ is the velocity dispersion of the lens DM halo.  Its uncertainty is estimated by
\begin{equation}
\delta D_{obs} = D_{obs} \sqrt{ \left(  \frac{\delta \theta_{E}}{\theta_E} \right)^2  +  4\left( \frac{\delta \sigma}{\sigma} \right)^2   }
\end{equation}
where $\delta \theta_E$ and $\delta \sigma$ are the uncertainty of $\theta_E$ and $\sigma$, respectively. Notice that $\delta D_{obs}$ does not consider correlation between  $\theta_E$ and $\sigma$. Following \cite{Amante:2019xao} we choice an absolute uncertainty on $\theta_E$ of $0.05$ for the data points without a reported uncertainty.

\subsection{Joint analysis}

In order to minimize the total $\chi^2$-function for each model given by
\begin{equation}
    \chi^2 = \chi^2_{OHD} + \chi^2_{SNIa} + \chi^2_{SLS}\,,
\end{equation}
we perform a Bayesian MCMC analysis based on \textit{emcee} module \cite{Emcee:2013}. After achieving a lower value than $1.1$ in the Gelman-Rubin criteria \cite{Gelman:1992} for each free model parameter to stop the n-burn phase, we obtain $5000$ chains with $250$ steps, each one to explore the confidence region taking into account a Gaussian prior on the Hubble constant $h$ and on DE density $\Omega_{m0}$ according to the Planck results \cite{Planck:2018}. Additionally, we select flat priors for $\lambda_0$ and $\lambda_1$ based on \cite{Hernandez-Almada:2019}. Table \ref{tab:priors} summarizes the priors used in the Bayesian analysis.

\begin{table}
\caption{Priors used in the MCMC analysis based on references  \cite{Planck:2018, Hernandez-Almada:2019}. Based on results in \cite{Hernandez-Almada:2019}, the parameter $n$ is fixed at $n=-2$.}
\centering
\begin{tabular}{|cc|}
\hline
Parameter      &  Prior                  \\
\hline
$h$            & Gauss$(0.6766, 0.0042)$ \\ [0.7ex]
$\Omega_{m0}$  & Gauss$(0.3111, 0.0056)$ \\ [0.7ex]
$\lambda_0$    & Flat in $[0,2]$         \\ [0.7ex]
$\lambda_1$    & Flat in $[0,2]$         \\ [0.7ex]
\hline
\end{tabular}
\label{tab:priors}
\end{table}

\section{Constraints and results} \label{Sec:ConsandRes}

\begin{figure}
  \centering
  \includegraphics[width=\linewidth]{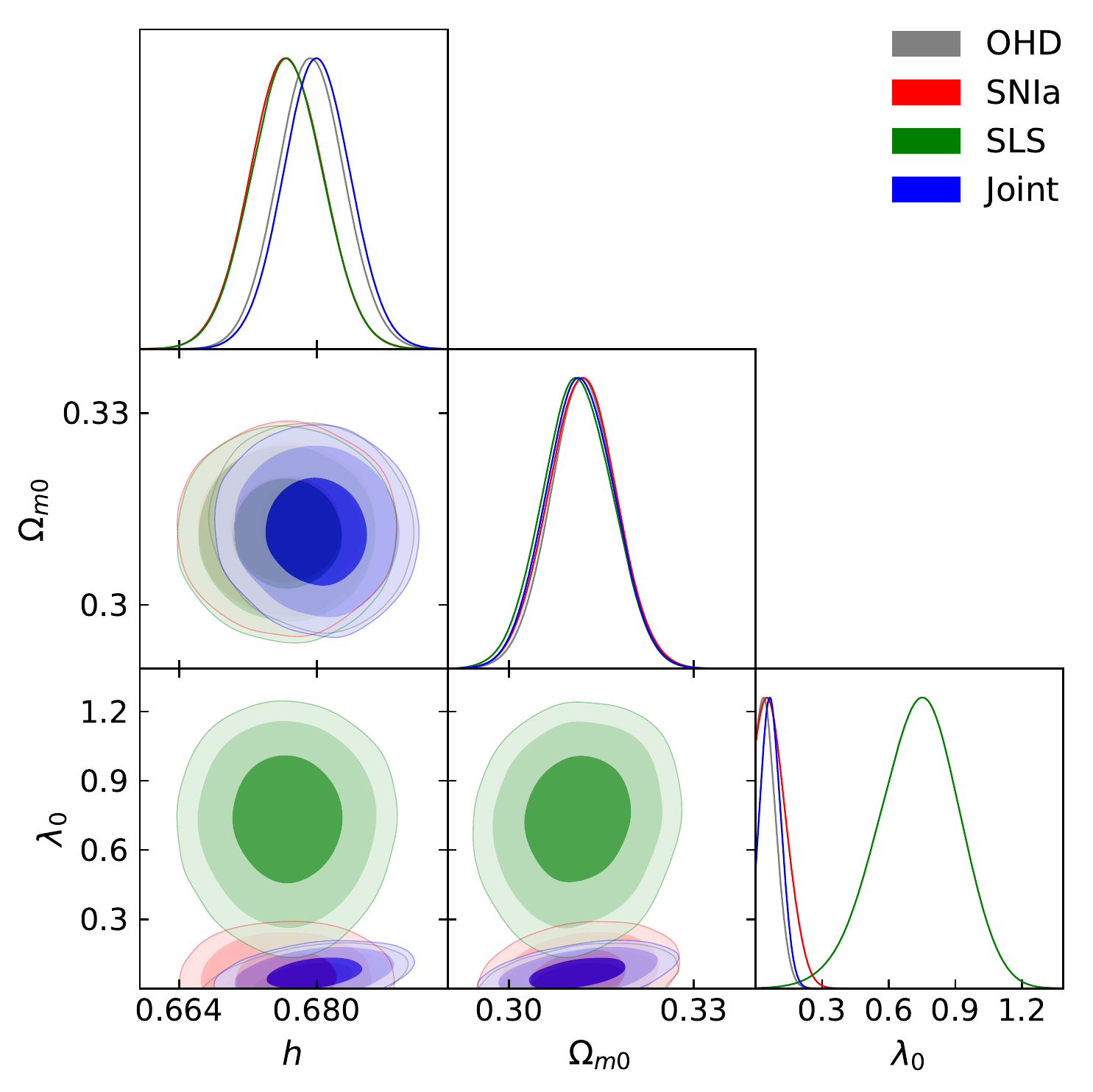} \\
  \includegraphics[width=\linewidth]{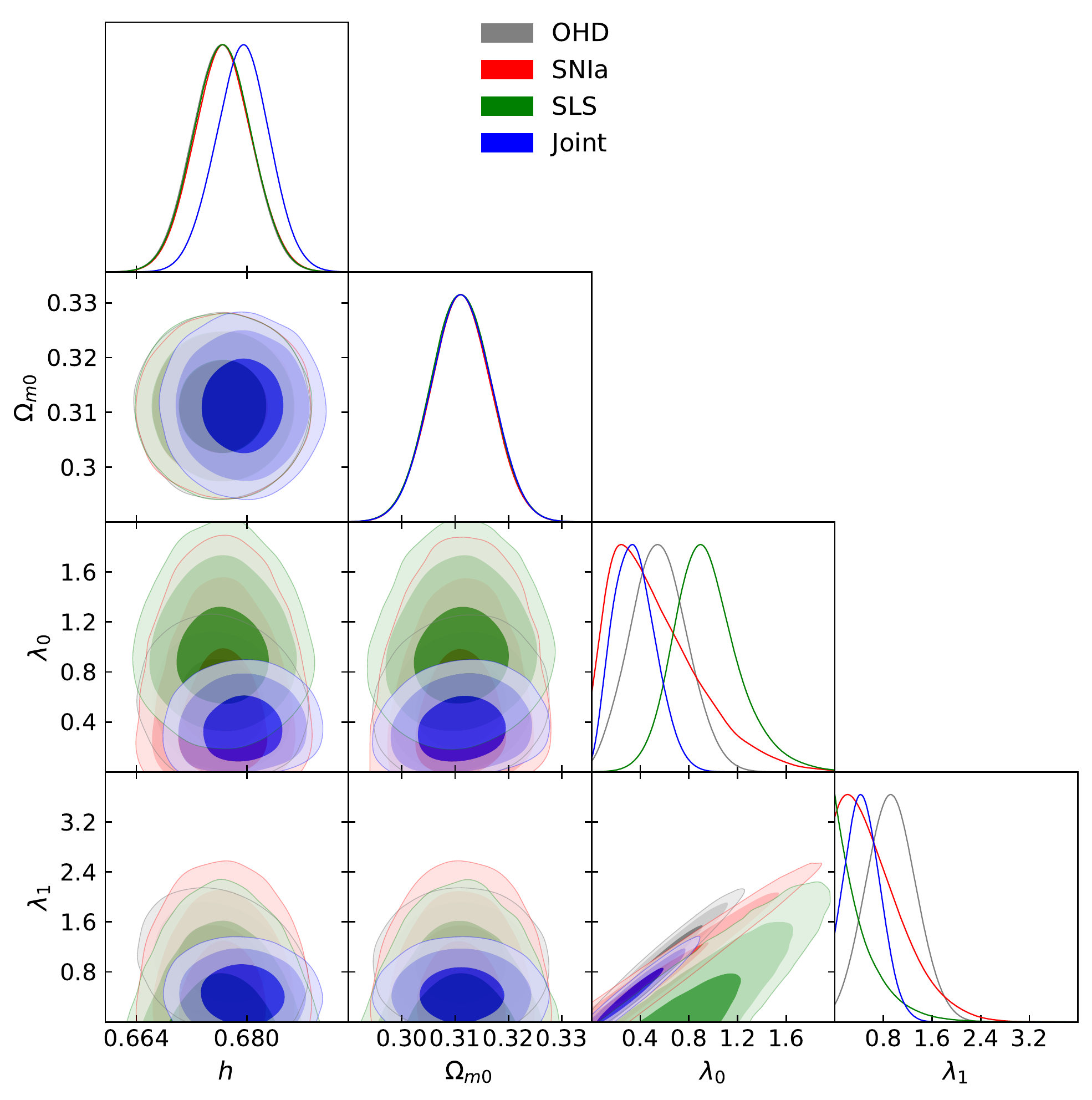} 
\caption{2D contour and 1D posterior distribution of the free parameters for the two fluids model when the viscosity coefficient is a constant (top panel) and polynomial (bottom panel) using the OHD, SNIa, SLS and OHD+SNIa+SLS (joint) data.}
\label{fig:contour_m2g}
\end{figure}

In this section we present and discuss our results obtained in the Bayesian analysis. 
Figure \ref{fig:contour_m2g} shows the constrained phase space of the parameters for CVM (top panel) and PVM (bottom panel), respectively at 68\% ($1\sigma$), 95\% ($2\sigma$), and 99.7\% ($3\sigma$) confidence level (CL). Table \ref{tab:bestfits} presents the mean fitting values obtained for both viscous models using OHD, SNIa, SLS and the joint analysis, respectively. The reported uncertainties correspond to $1\sigma$ CL. We find good agreement to data according to the chi-square value of the models, being the worse fit of them to SLS data with $\chi^2=602.3$ ($603.6$) for the CVM (PVM).
It is worth mentioning that CVM and PVM are better approximations to LCDM because consider a DE component with EoS $w=-1$ and a viscous matter component. By setting the $\lambda=0$ (or $\xi=0$), we recover the LCDM with DE and DM components, which the relativistic species are negligible. In other words, we are studying as a first approximation the consequences of the viscosity effects in the LCDM model under the Eckart formalism.
In this sense, it is convenient to compare statistically our models with LCDM model. Instead of using $\chi^2$ criteria, it is more convenient to use others such as the Akaike information criterion (AIC) \cite{AIC:1974, Sugiura:1978} and Bayesian information criterion (BIC) \cite{schwarz1978} because they allow to compare models with different degree of freedom. They are defined as ${\rm AIC}= \chi^2+2k$ and ${\rm BIC}=\chi^2+k\log(N)$ where $k$ is the number of free parameters, and $N$ is the size of the data sample. In these approaches, the model with lowest values of AIC (BIC) is preferred by data. To contrast with LCDM model, we consider the values $\Omega_{m0}=0.3111$, $h=0.6766$ reported by \cite{Planck:2018}, and $\mathcal{M}=-19.408$ to obtain AIC and BIC values 
${\rm AIC}^{LCDM} = 29.4,1032.5, 1674.4$ and 
${\rm BIC}^{LCDM} = 33.3, 1047.4, 1689.9$ 
using OHD, SNIa, and joint analysis. It is worth to mention that we have not estimated BIC and AIC for LCDM using SLS data because this sample can not constrain either $h$ or $\Omega_{m0}$. Following the rules described in \cite{Almada:2020}, we discuss our results. 
For CVM, we obtain 
$\Delta {\rm AIC}^c ={\rm AIC}^c-{\rm AIC}^{LCDM}= 2.5, 2.5, 17.7$, 
$\Delta {\rm BIC}^c ={\rm BIC}^c-{\rm BIC}^{LCDM}= 4.4, 7.5, 22.9$. 
Then, according to AIC values, we have that the CVM and LCDM model are equally supported considering OHD and SNIa data and we do not have support for the CVM over LCDM taking into account the joint analysis. In BIC, we have a weak evidence against the CVM for OHD, a stronger evidence against CVM on SNIa analysis and, an even stronger evidence against this viscous model over LCDM on the joint analysis.
For the PVM, 
we obtain 
$\Delta {\rm AIC}^p  = -0.5, 22.0,  4.4$ and 
$\Delta {\rm BIC}^p   =  3.3, 31.9, 14.7$. We observe equal support for both models, PVM and LCDM for OHD, and although, we have even strong evidence against PVM according to BIC, we obtain similar support for both models in the AIC.
Additionally, we find that the CVM is equally supported than  $\omega$CDM, the Chevallier-Polarski-Linder (CPL), and the Jassal-Bagla-Padmanabhan (JBP) models \citep{Amante:2019xao} and no evidence against as well. In contrast, we find a slightly less support for PVM than previous mentioned parameterizations, and a strong evidence against PVM over them.
 
\begin{table*}
\caption{Best fitting values of the free model parameters.}
\centering
\begin{tabular}{|ccccccccc|}
\hline
Sample     &    $\chi^2$     &  $h$ & $\Omega_{m0}$ & $\lambda_0$ & $\lambda_1$ &  $\mathcal{M}$   & AIC & BIC          \\
\hline
\multicolumn{9}{|c|}{$\xi_0={\rm Constant}$} \\ 
 OHD   & $25.9$ & $0.679^{+0.004}_{-0.004}$ & $0.312^{+0.005}_{-0.005}$ & $0.053^{+0.047}_{-0.035}$ & -- & -- & $31.9$ & $37.7$ \\ [0.7ex]
SNIa  & $1027.1$ & $0.676^{+0.004}_{-0.004}$ & $0.312^{+0.005}_{-0.005}$ & $0.080^{+0.071}_{-0.072}$ & -- & $-19.400^{+0.016}_{- 0.016}$  & $1035.1$ & $1054.9$ \\ [0.7ex]
SLS  & $602.3$ & $0.677^{+0.004}_{-0.004}$ & $0.311^{+0.006}_{-0.006}$ & $0.737^{+0.175}_{-0.188}$ & -- & --   & $606.3$ & $612.9$\\ [0.7ex]
 Joint & $1684.1$ & $0.680^{+0.004}_{-0.004}$ & $0.311^{+0.006}_{-0.005}$ & $0.071^{+0.047}_{-0.040}$ & -- & $-19.400^{+0.012}_{- 0.012}$  & $1692.1$ & $1712.8$ \\ [0.7ex]
\hline
\multicolumn{9}{|c|}{ $\xi_0={\rm Polynomial}$ } \\ [0.7ex]
OHD  & $20.9$   & $0.676^{+0.004}_{-0.004}$ & $0.311^{+0.006}_{-0.006}$ & $0.551^{+0.237}_{-0.228}$ & $0.929^{+0.412}_{-0.401}$ & --  &  $28.9$ & $36.6$ \\ [0.7ex]
SNIa & $1044.5$ & $0.676^{+0.004}_{-0.004}$ & $0.311^{+0.006}_{-0.006}$ & $0.461^{+0.441}_{-0.280}$ & $0.580^{+0.620}_{-0.395}$ & $-19.400^{+0.019}_{- 0.019}$  &  $1054.5$ & $1079.3$ \\ [0.7ex]
SLS  & $603.6$  & $0.676^{+0.004}_{-0.004}$ & $0.311^{+0.006}_{-0.006}$ & $0.927^{+0.284}_{-0.235}$ & $0.312^{+0.484}_{-0.231}$ & --  & $609.6$ & $619.6$ \\ [0.7ex]
Joint & $1668.8$ & $0.679^{+0.004}_{-0.004}$ & $0.311^{+0.006}_{-0.006}$ & $0.347^{+0.183}_{-0.164}$ & $0.465^{+0.301}_{-0.263}$& $-19.400^{+0.014}_{- 0.014}$  & $1678.8$ & $1704.7$  \\ [0.7ex]
\hline
\end{tabular}
\label{tab:bestfits}
\end{table*}

\begin{figure*}
  \centering
  \includegraphics[width=.48\linewidth]{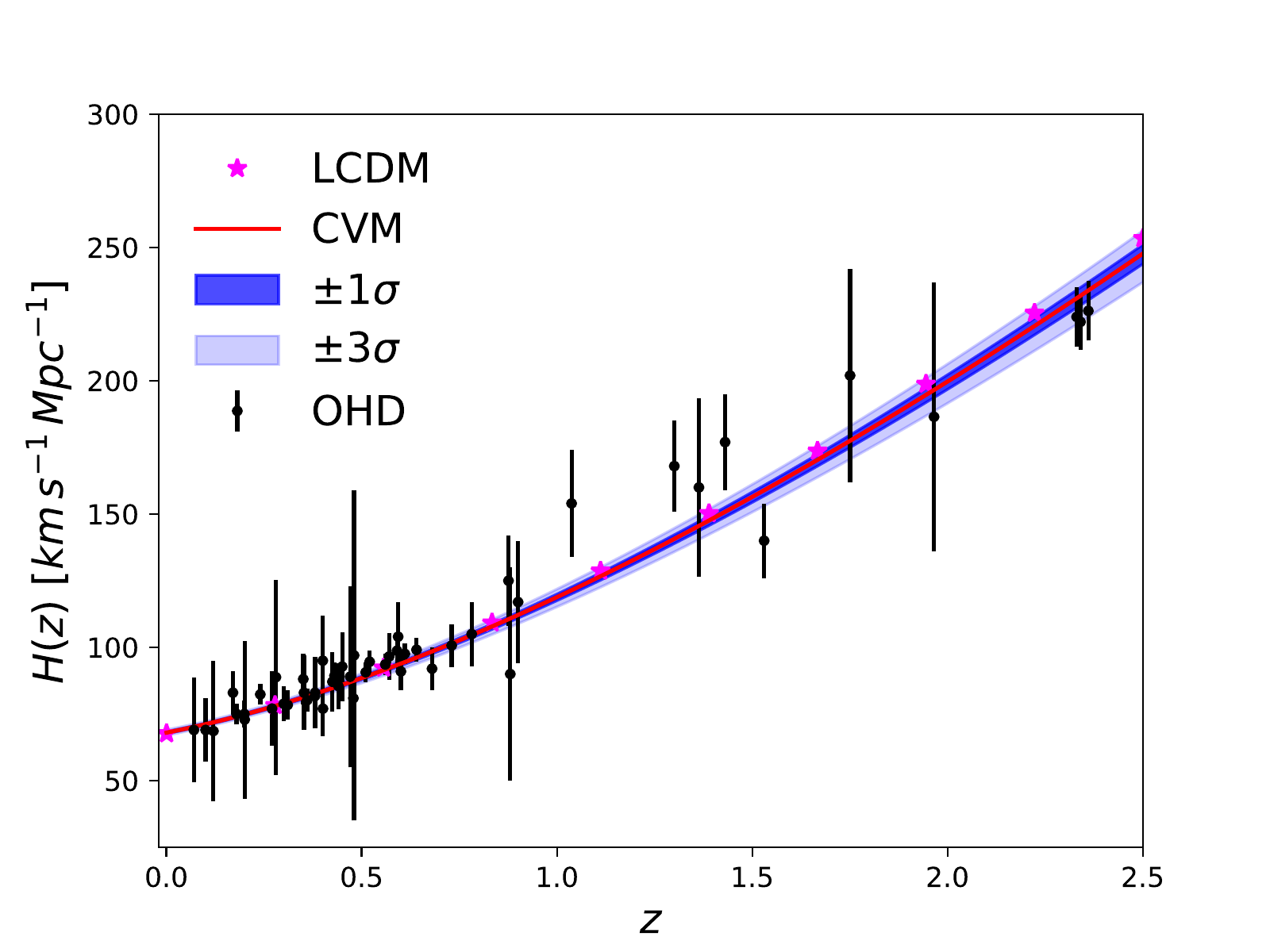} 
  \includegraphics[width=.48\linewidth]{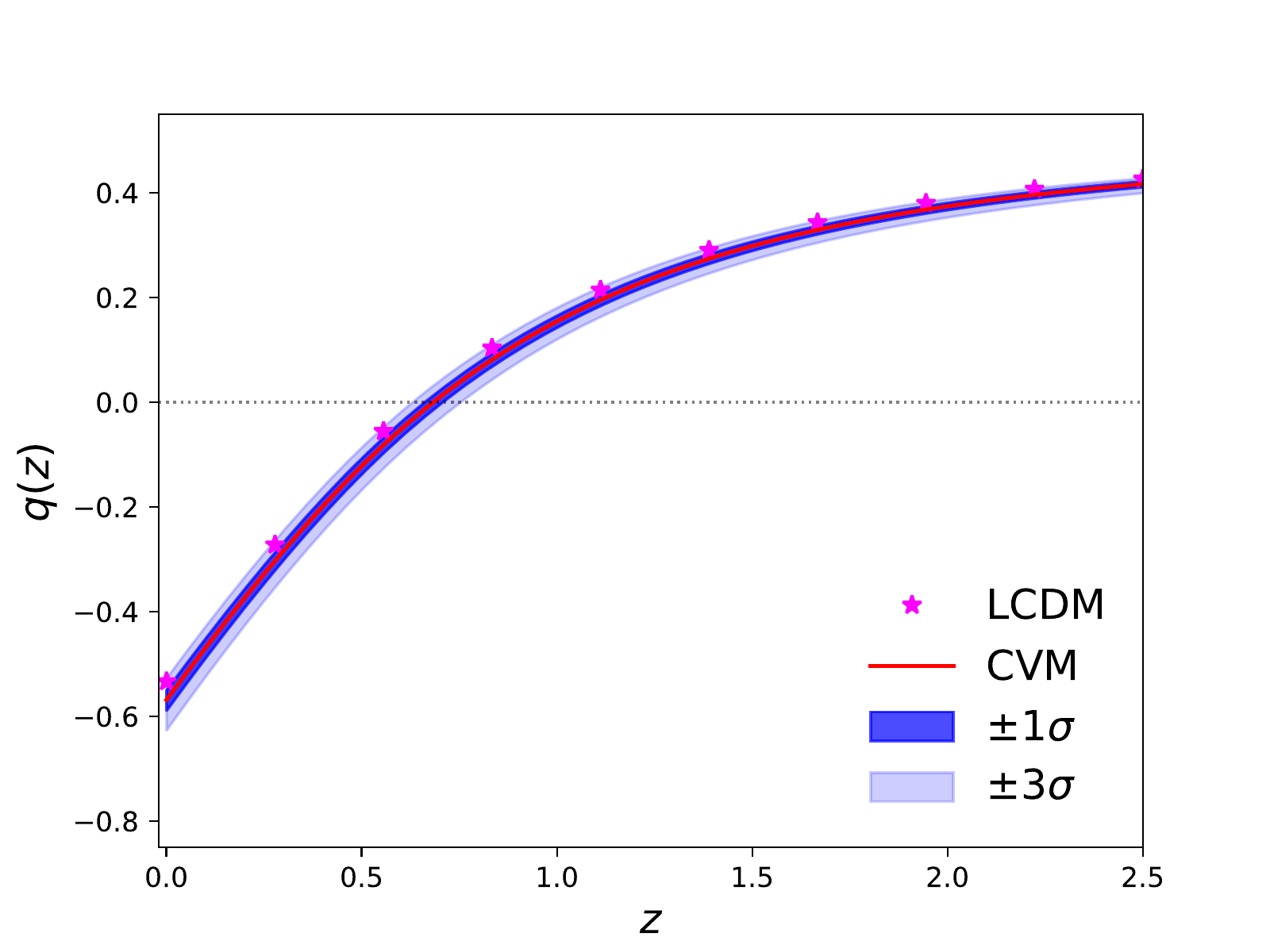} \\
  \includegraphics[width=.48\linewidth]{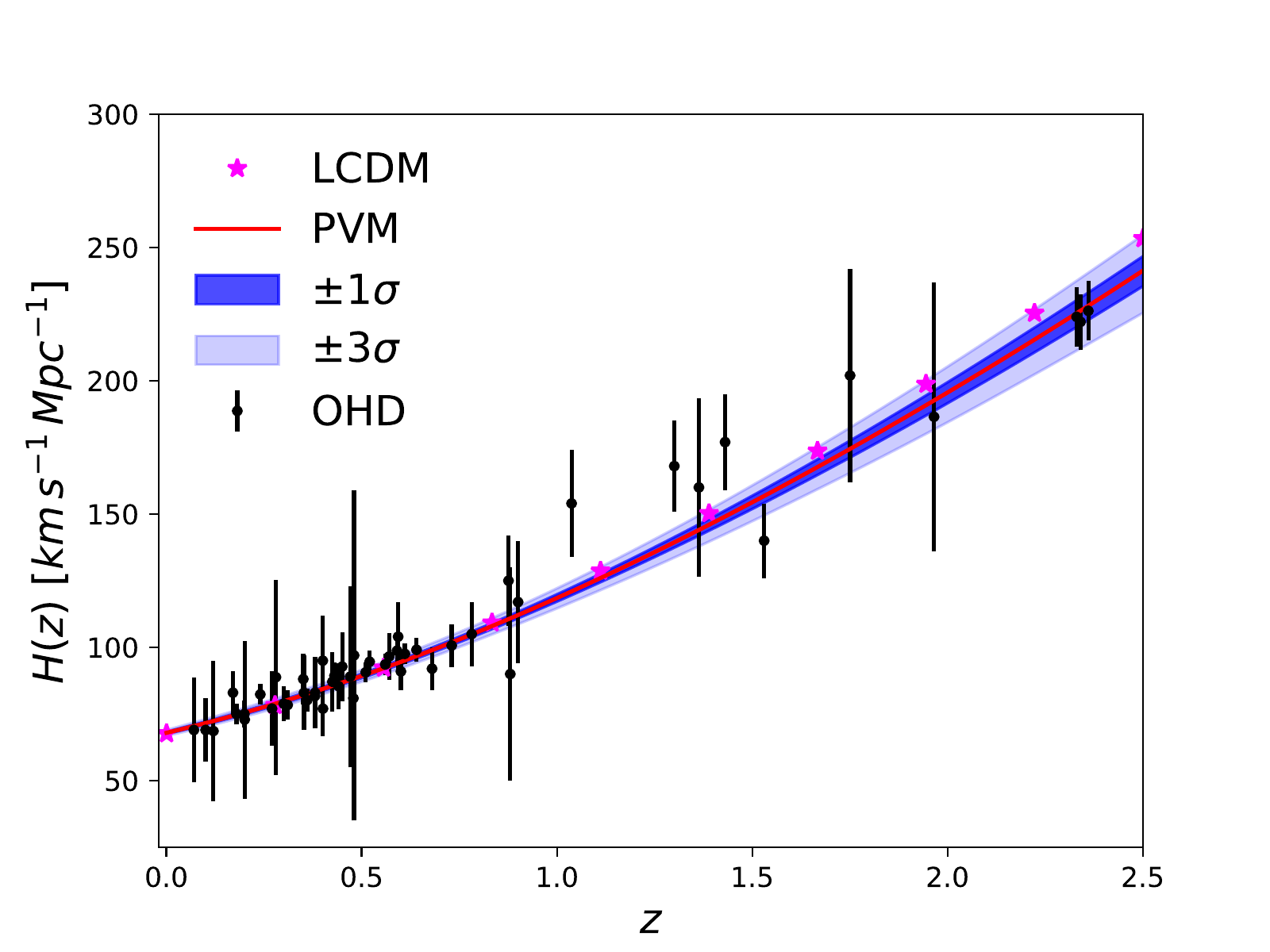} 
  \includegraphics[width=.48\linewidth]{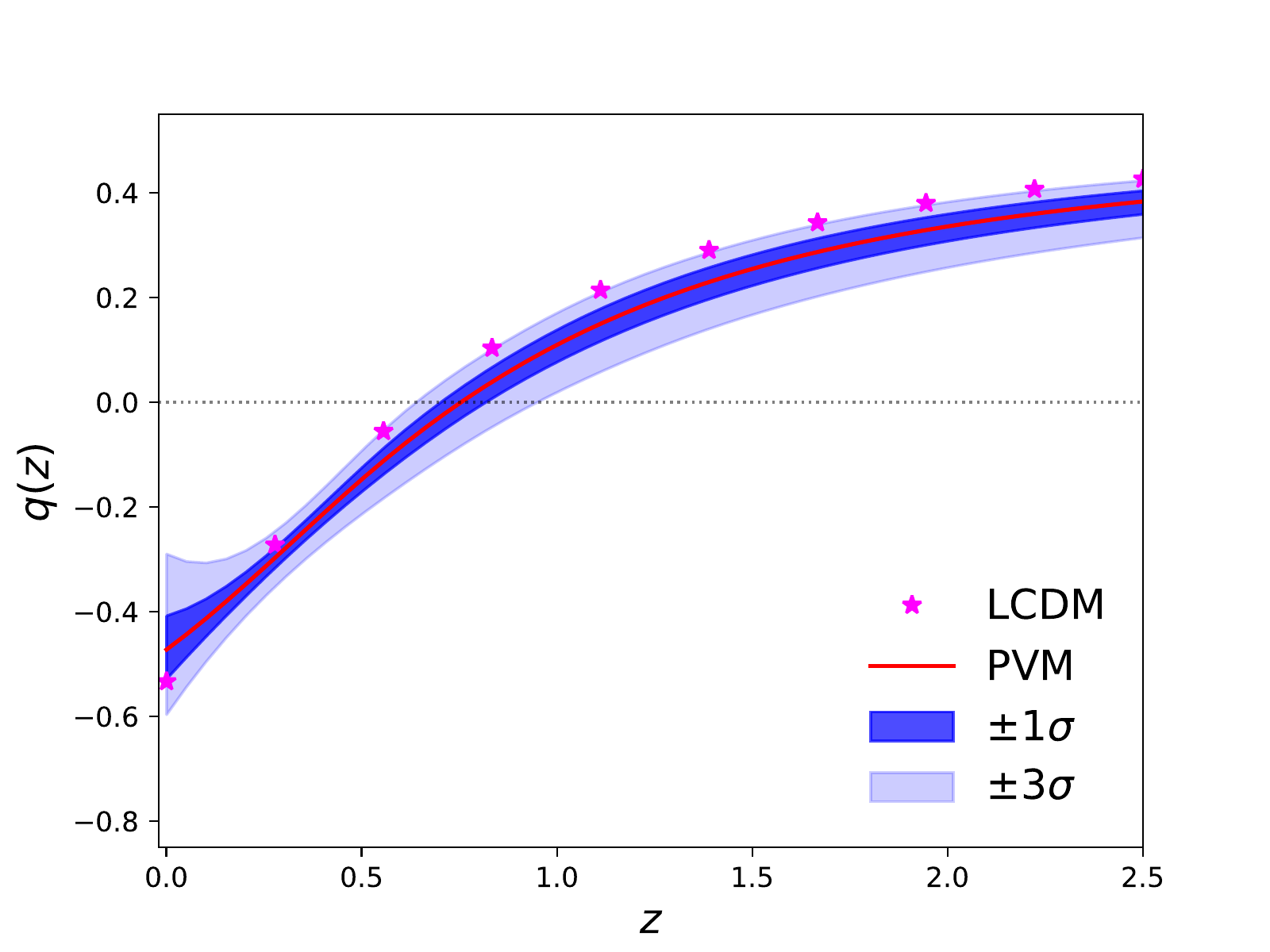} \\
\caption{Left panel: Best fits over OHD for the constant (polynomial) model at the top (bottom) panel. Right panel:  deceleration parameter. The darker (lighter) bands correspond the uncertainty at $1\sigma$ ($3\sigma$) CL. The (magenta) star marker represents the reconstruction of the LCDM model.}
\label{fig:fitOHD}
\end{figure*}

\begin{figure*}
  \centering
  \includegraphics[width=.32\linewidth]{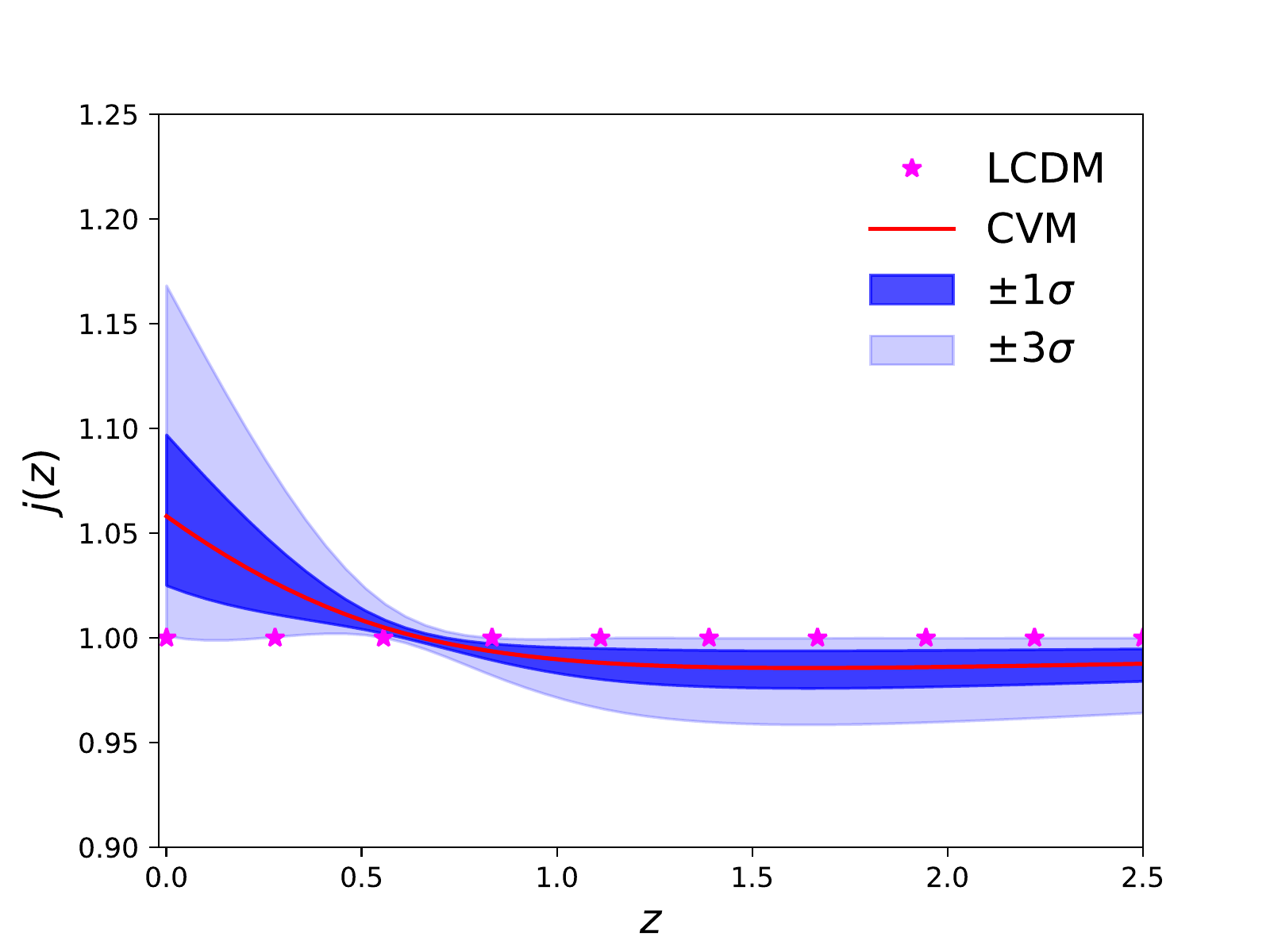}   
  \includegraphics[width=.32\linewidth]{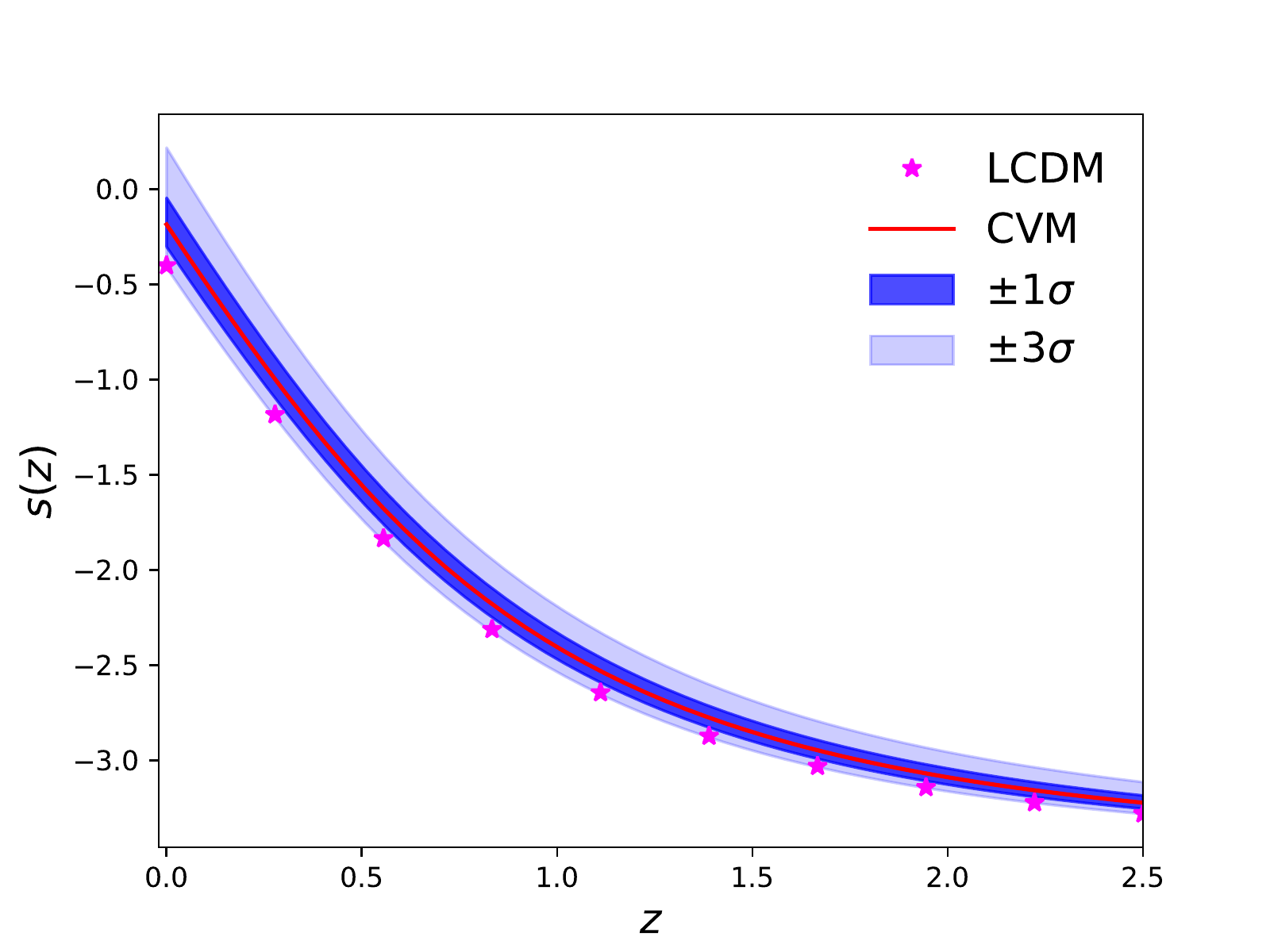}    
  \includegraphics[width=.32\linewidth]{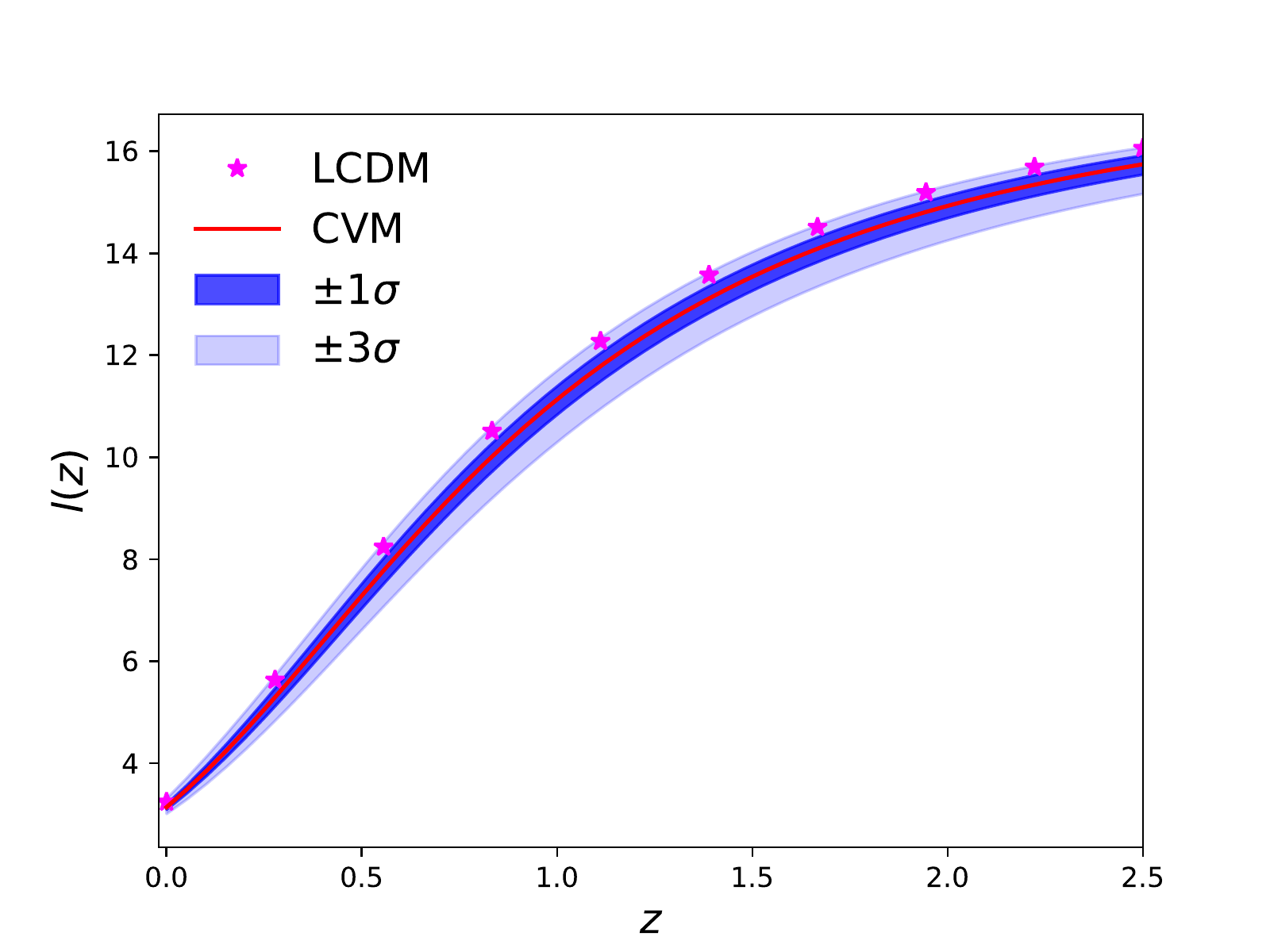} \\
  \includegraphics[width=.32\linewidth]{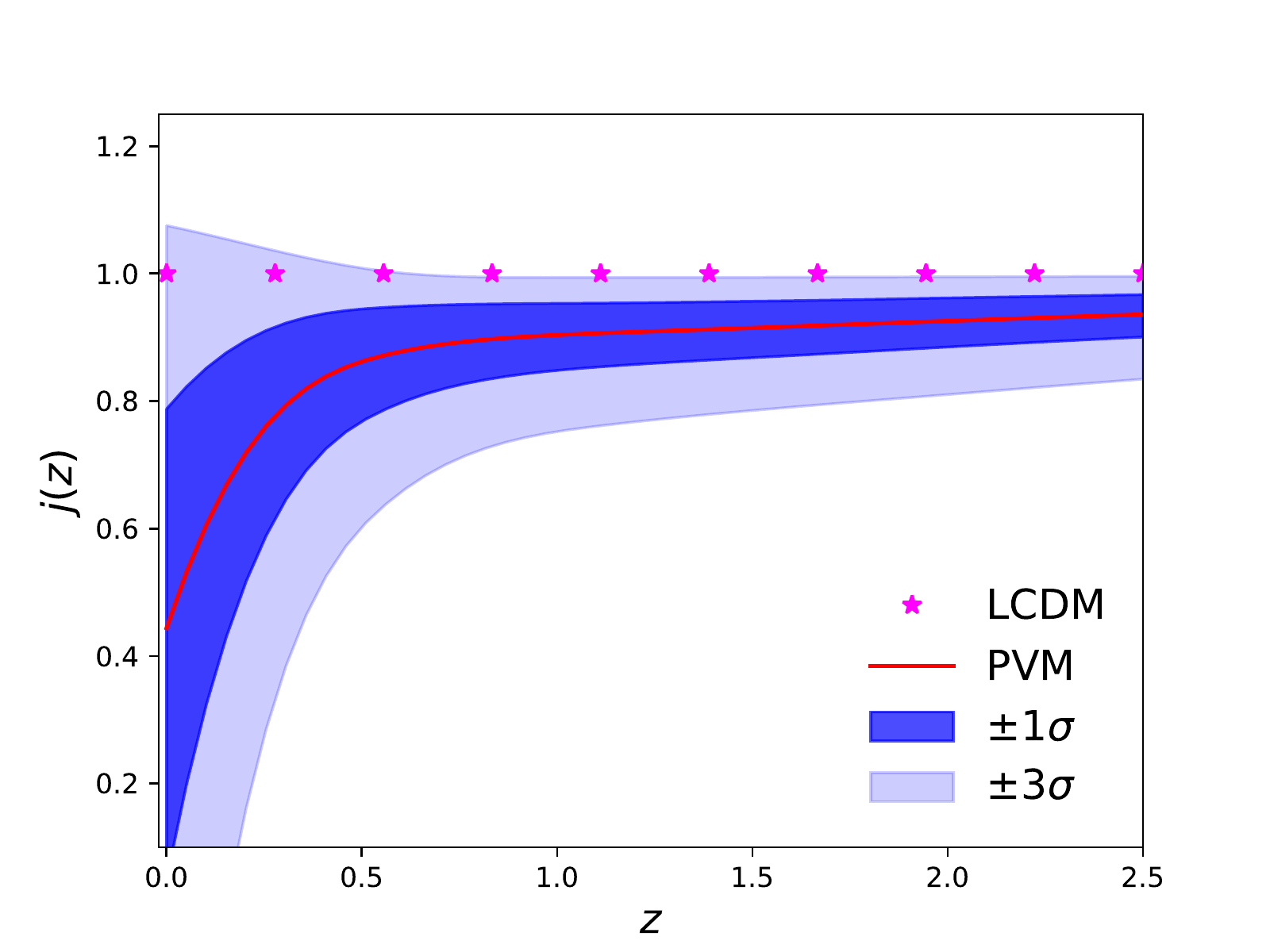}
  \includegraphics[width=.32\linewidth]{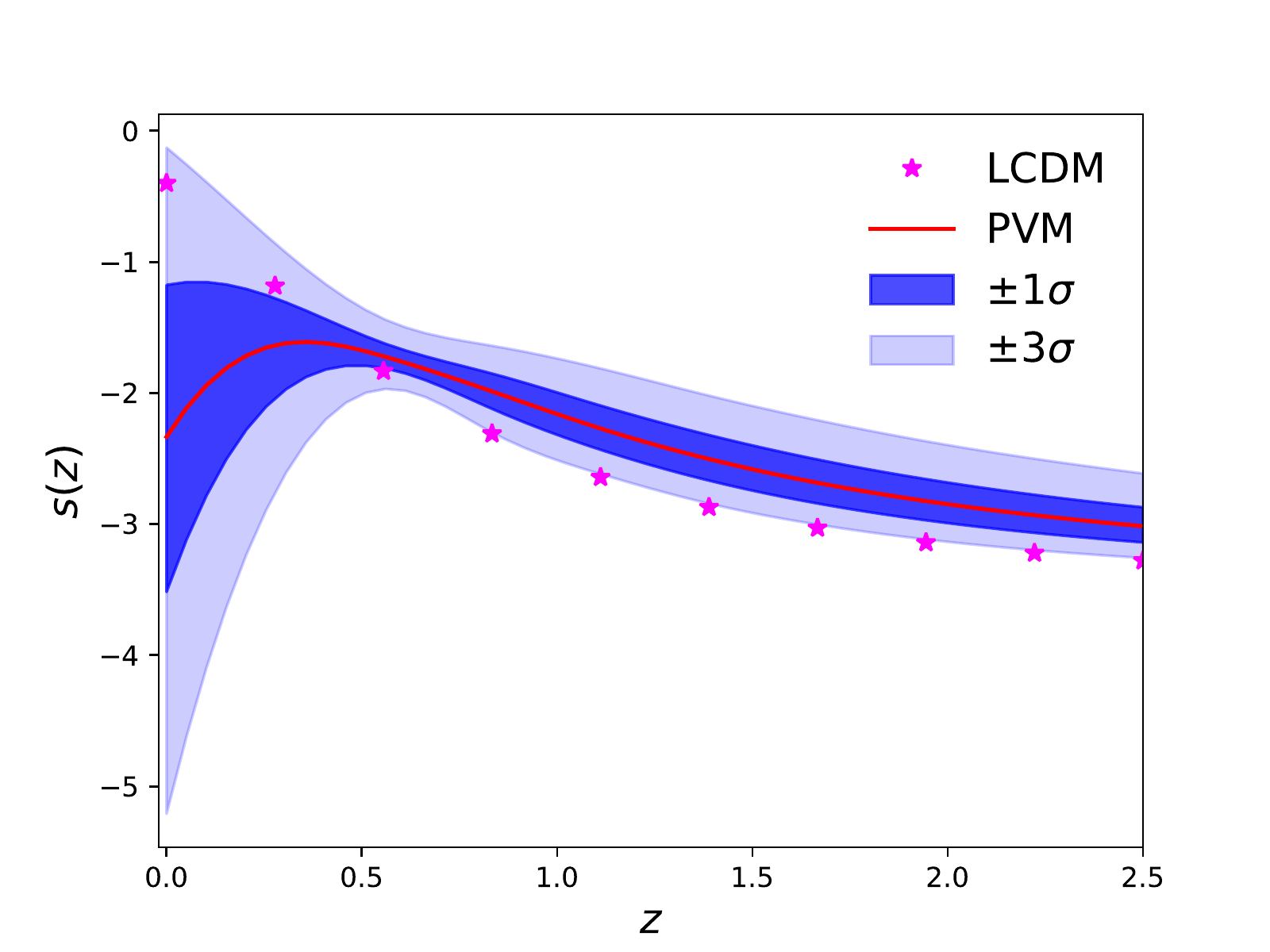}
  \includegraphics[width=.32\linewidth]{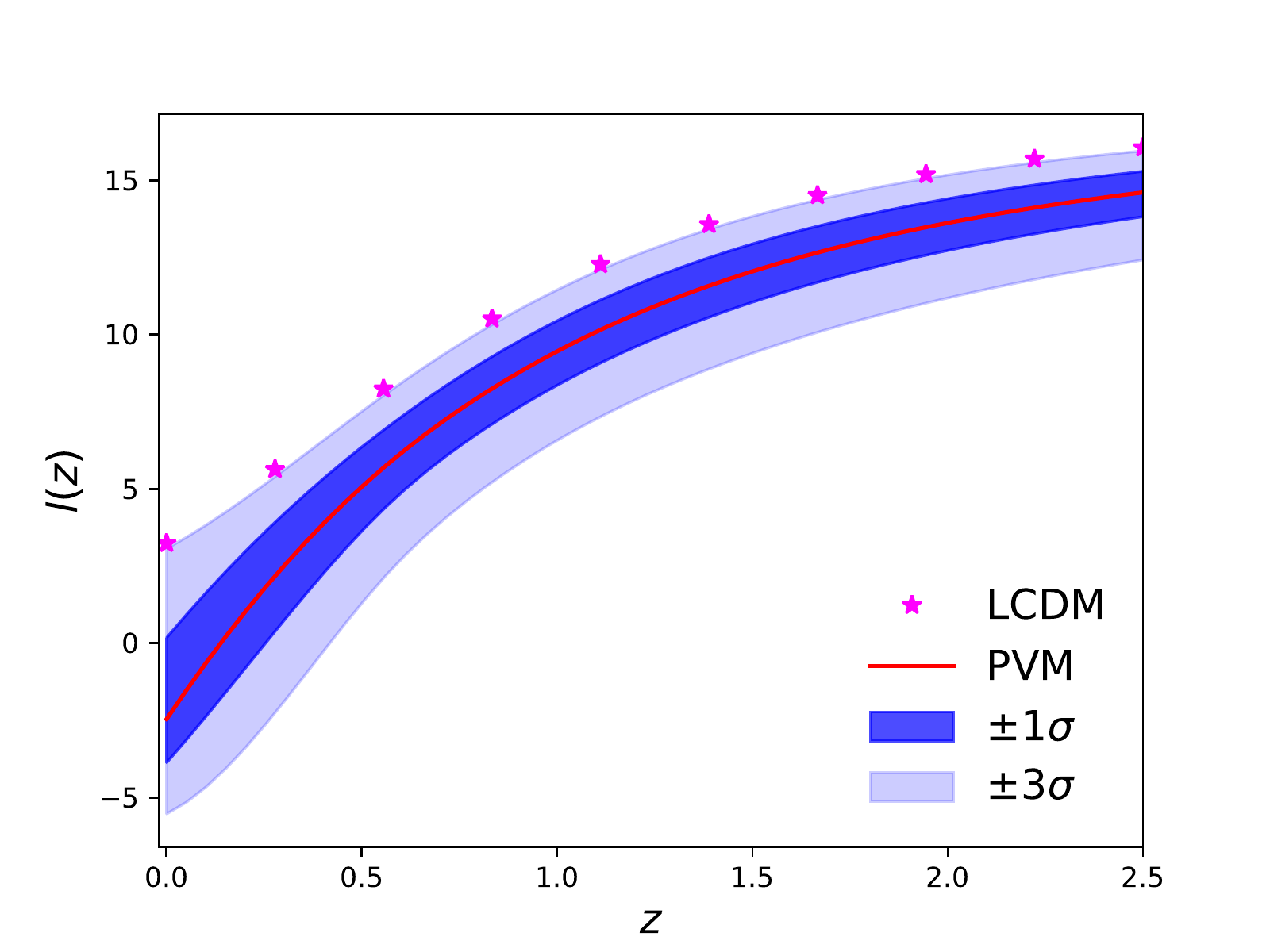}\\ 
\caption{Cosmographic parameter reconstruction. From left to right, it is the jerk (j), snap (s), and lerk (l) parameters. At top panel is the cosmographic variables for the constant model and at the bottom is the ones for the polynomial model.}
\label{fig:jsl}
\end{figure*}

We reconstruct the cosmographic parameters for both viscous models, CVM and PVM, in the redshift region $0<z<2.5$. Figure \ref{fig:fitOHD} displays the Hubble and deceleration parameters for CVM (top panel) and PVM (bottom panel) respectively. We find a good agreement with LCDM (star markers) within $3\sigma$ CL along the mentioned range. 
Additionally, we calculate the corresponding deceleration-acceleration transition redshift values, obtaining 
$z_t = 0.683^{+0.025}_{-0.022}$ and
$z_t = 0.755^{+0.063}_{-0.051}$. When we compare them with LCDM value ($z_t^{LCDM}=0.642^{+0.014}_{-0.014}$), we obtain a deviation of $1.9\sigma$ and $2.2\sigma$. Figure \ref{fig:jsl} shows the reconstruction of the high order cosmographic parameters ($j$, $s$, $l$) and their uncertainties at $1\sigma$ and $3\sigma$. The jerk parameter gives us information about the dynamics of the DE EoS, corresponding to $j=1$ to $w=-1$ for the DE component. In this sense, although our models are consistent up to $3\sigma$ with LCDM, our best fitting values indicate an effective dynamical DE EoS. In other words, we observe a deviation to LCDM in the jerk parameter due to the viscosity contributions. 

On the other hand, it is well known that the snap and lerk parameters and higher cosmographic parameters does not have a well established physical meaning, however they are an important part of the Taylor series of the Hubble parameter in cosmography, giving us more precision in the preferred model by observations. Notoriously, the viscosity generates important differences (for snap and lerk) in the PVM case in comparison with LCDM, mainly at lower redshift, which in turn means that dark energy is dynamic and not constant in this kind of models. This is consistent with cosmographic studies where the LCDM is not the preferred model (see Refs. \cite{Demianski:2012,Aviles:2012ay,Aviles:2016wel,Zhang:2016urt} for details).

\begin{figure*}
  \centering
  \includegraphics[width=.45\linewidth]{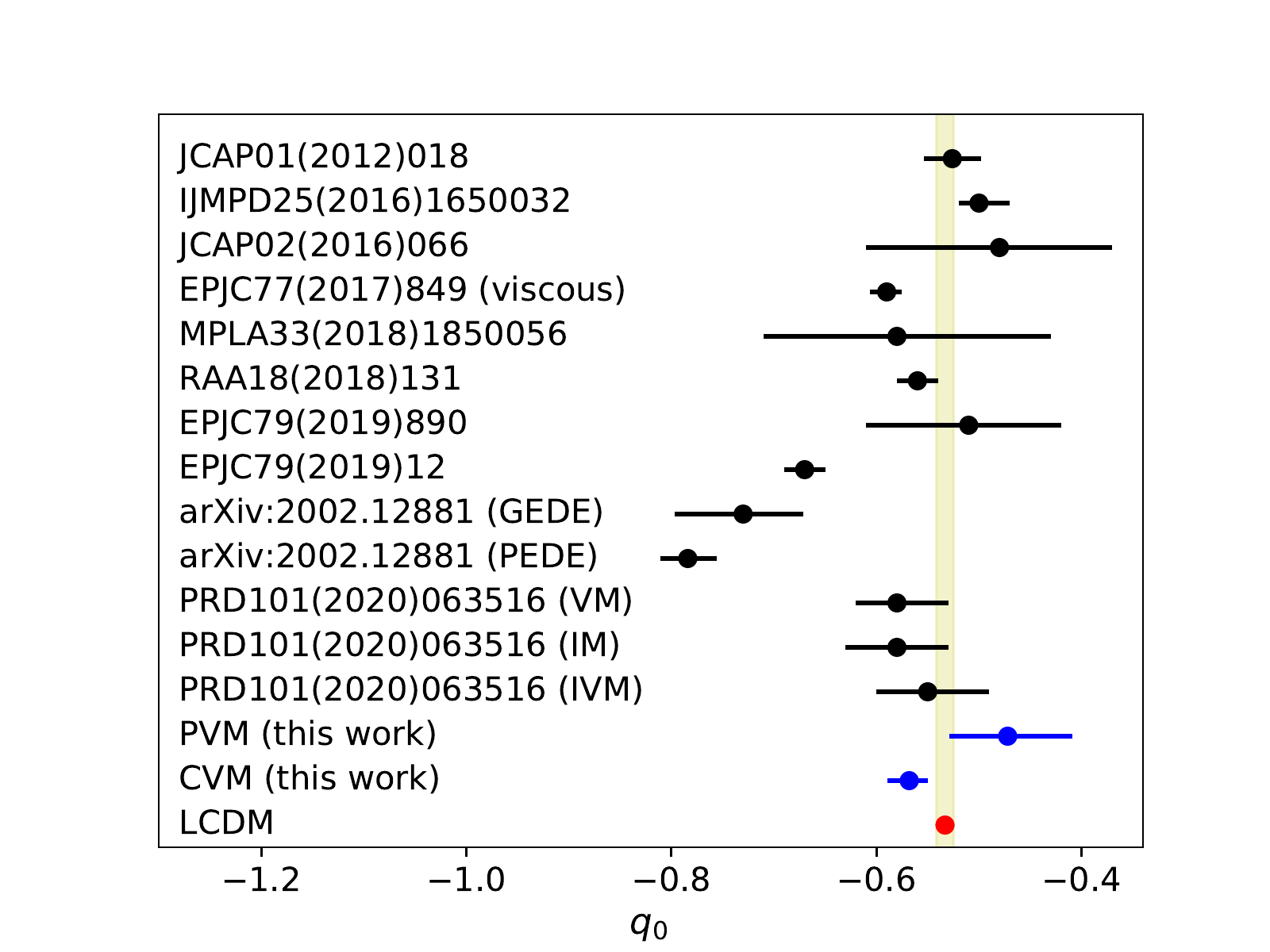} 
  \includegraphics[width=.45\linewidth]{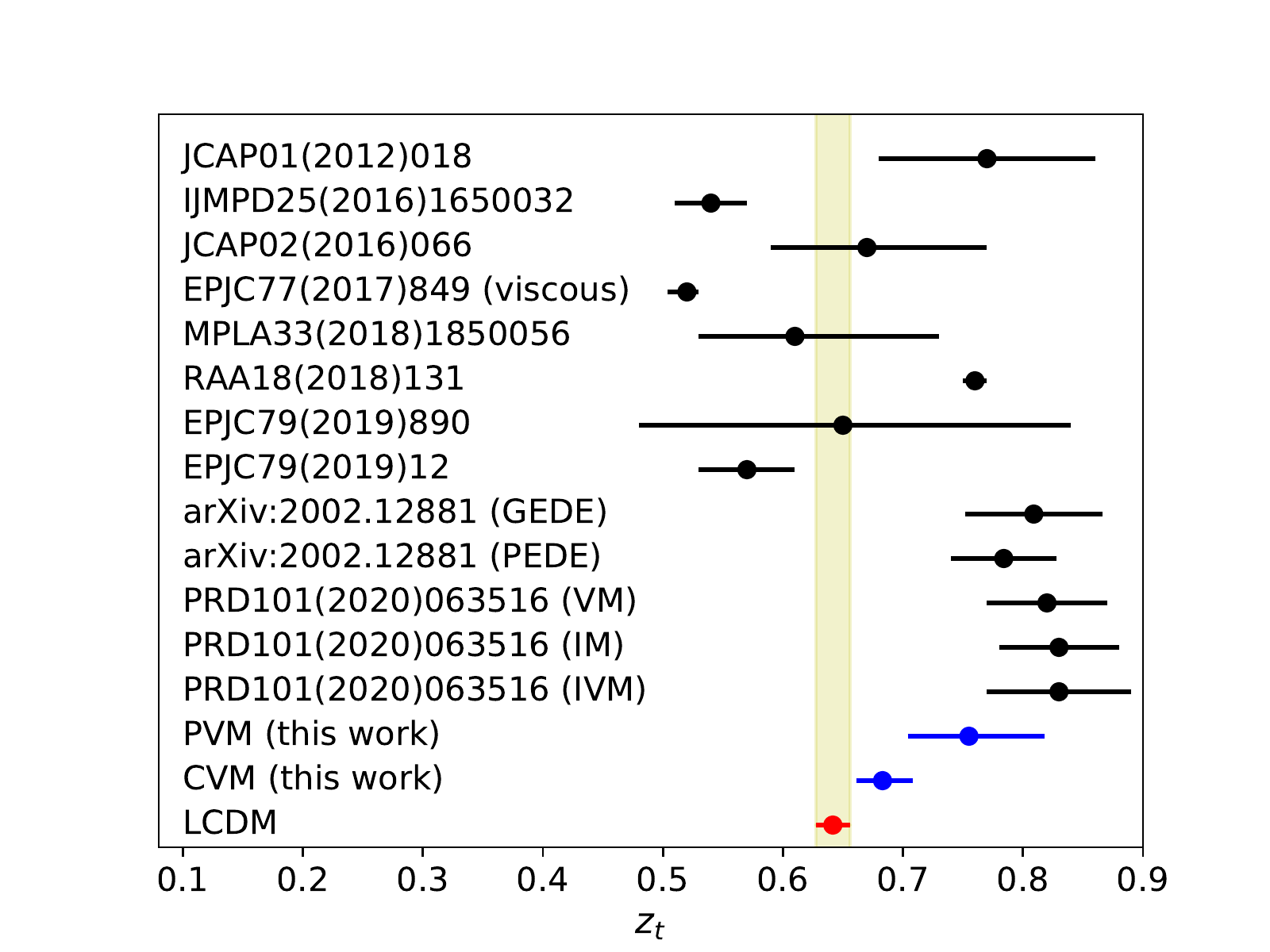}    
\caption{Comparison of the $q_0$ and $z_t$. The vertical band represents the uncertainty at $1\sigma$ of the LCDM model.}
\label{fig:j0zt}
\end{figure*}

Furthermore, we report the cosmographic parameters at current epochs,  
$q_0   = -0.568^{+0.018}_{-0.021}$
$j_0   = 1.058^{+0.039}_{-0.033}$
$s_0   = -0.184^{+0.141}_{-0.117}$
$l_0   = 3.139^{+0.056}_{-0.051}$
and 
$q_0   = -0.472^{+0.064}_{-0.056}$,
$j_0   = 0.444^{+0.344}_{-0.394}$,
$s_0   = -2.334^{+1.159}_{-1.184}$,
$l_0   = -2.460^{+2.636}_{-1.412}$
for the constant and polynomial form of the viscosity, respectively. We have a deviation of about $1.9\sigma$ ($1.1\sigma$) between CVM (PVM) value and the LCDM one ($q_0^{LCDM}=-0.533^{+0.008}_{-0.008}$).
Furthermore, the values of $q_0$ are consistent with those reported in the literature \cite{Hernandez-Almada:2019} within $2.2\sigma$ when hyperbolic functions are considered. Additionally, the authors \cite{Hernandez-Almada:2019} also reports $q_0=-0.680^{+0.085}_{-0.102}$ for a Universe filled by a single non-perfect fluid with polynomial viscosity, achieving a deviation of $5.3\sigma$ ($3.7\sigma$) relative to CVM (PVM). It is interesting to observe that PVM is a generalization of this polynomial single fluid model because PVM includes a DE component but has a better approximation to LCDM. In summary, Fig. \ref{fig:j0zt} compares the $q_0$ and $z_t$ obtained for several models reported in the literature  (see these Refs. \cite{Nair_2012, Sudipta:2016, Santos_2016, Mohan:2017, Abdulla-MPLA:2018, Das_2018, Hernandez-Almada:2018osh, Garza:2019, Hernandez-Almada:2019, Almada:2020, Almada-GEDE:2020} for more details about the models and cosmological data used). The vertical band represents $\pm1\sigma$ around the central value of LCDM. It is interesting to observe that models with interactions and viscosity have a deceleration-acceleration transition earlier than LCDM and most of the models are in good agreement with the expected value $q_0$ for LCDM. Additionally, we estimate the effective EoS at current times, obtaining values of $w_{eff0} = -0.712^{+0.012}_{-0.014}$ and $w_{eff0}=-0.648^{+0.043}_{-0.038}$ for CVM and PVM respectively. These values have a deviation from one reported in \cite{Hernandez-Almada:2018osh} (Chaplygin-like model) of $3.95\sigma$ and $3.36\sigma$ respectively. Furthermore, we find a deviation of $5.66\sigma$ and $1.43\sigma$ to the value presented in \cite{NCruz:2019} (dissipative dark fluid model), and good agreement (up to $2.1\sigma$) with the one reported in \cite{Mohan:2017}.

\begin{figure}
  \centering
  \includegraphics[width=\linewidth]{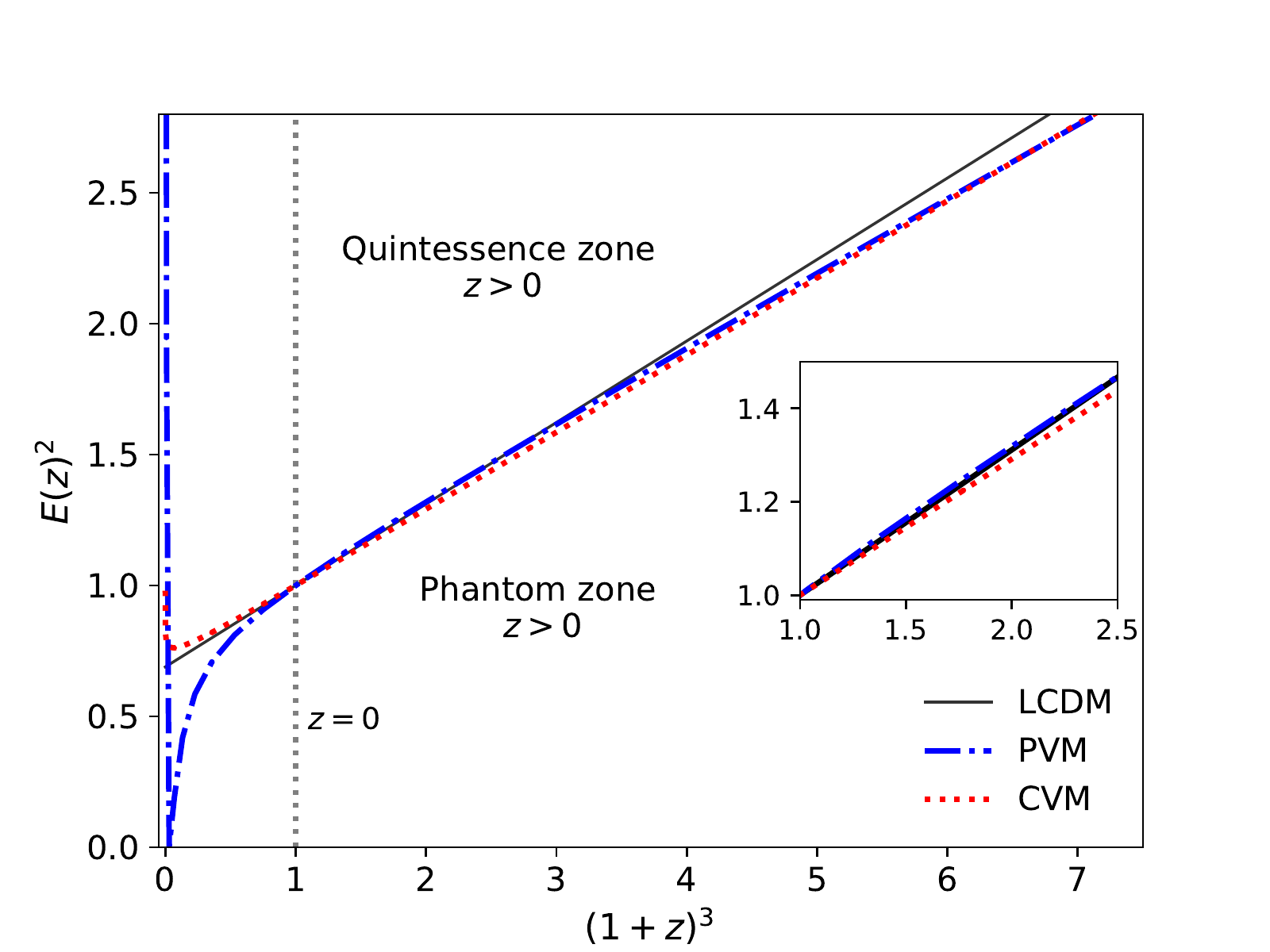} 
\caption{Behavior of $E(z)^2$ over $(1+z)^3$. Solid black line represents LCDM model, blue dot-dashed line is the PVM, and red dotted line is the CVM. For the PVM, we observe a slightly transition from phantom to quintessence behaviour in the region $1<z<2.5$. Vertical dotted line represents the current Universe ($z=0$).}
\label{fig:E2u3}
\end{figure}

Figure \ref{fig:E2u3} displays the evolution of $E(z)^2$ as function of $(1+z)^3$ for both viscous models. In this diagram, the black line represents the LCDM, and the region $z>0$ over LCDM corresponds to a quintessence behavior and the region $z>0$ below LCDM line to a phantom one. In this vein, it is interesting to observe that CVM behaves as phantom DE for any time, presenting a divergence in the future. Although, PVM also presents a singularity in the future ($z=-1$), PVM presents a transition from phantom to quintessence around $z\approx 2.2$, conserving such behavior close to $z=-1$, before to finish in a Big Rip. It is worth mentioning that a Big Rip at the future is a typical final state presented in the viscous models (see for instance \cite{Brevik:2011mm, Normann:2016}).

On the other hand, we estimate the correlation between cosmographic parameters, $z_t$, and those proper of the models through the formulae $\rm{corr}(x,y)=\rm{Cov}(x,y)/\sigma_x\sigma_y$ where $\rm{Cov}(x,y)$ is the covariance coefficient between $x$ and $y$, and $\sigma_x$ ($\sigma_y$) is the standard deviation of $x$ ($y$).  We find a strong correlation (abs(corr)$>0.7$) between those related to the viscosity ($\lambda_0$ and $\lambda_1$) and $q_0$, and between $z_t$. In other words, the viscosity terms allow us to modify both physical quantities that characterize the Universe. Additionally, we also find a negative strong correlation between ($\lambda_0,\lambda_1$) and the high order cosmographic parameters ($j_0,s_0, l_0$) at current epochs.

\begin{figure}
  \centering
  \includegraphics[width=\linewidth]{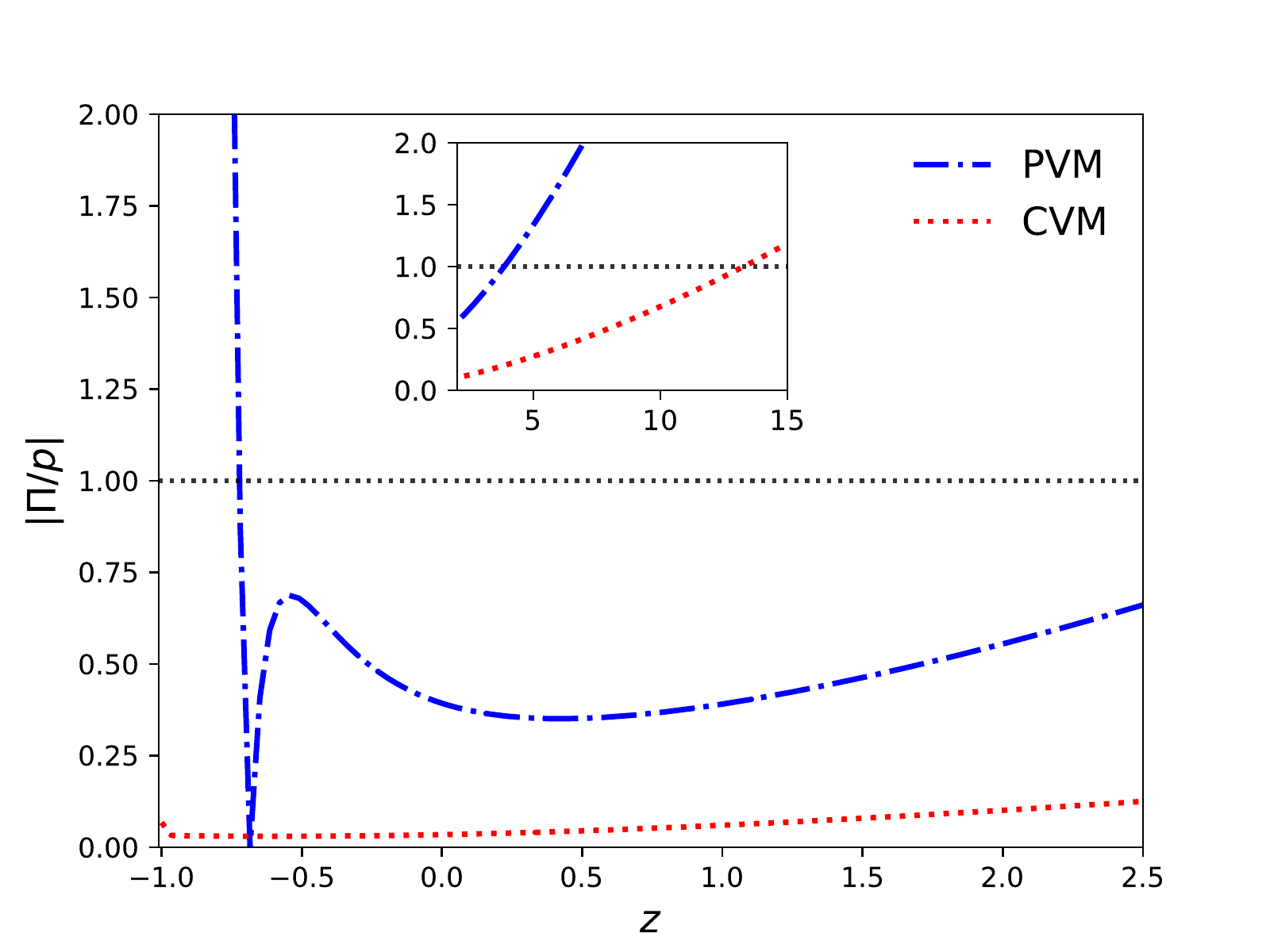} 
\caption{Evolution of the quotient $|\Pi/p|$ over $z$, in the redshift region $-1<z<2.5$ and $2.5<z<15$ for the inner panel. The PVM and CVM are represented by blue dot-dashed line and red dotted line respectively.}
\label{fig:Pipvsz}
\end{figure}

Finally, although it has been shown (see for instance \cite{Mohan:2017,Hernandez-Almada:2019}) that with an unique dissipative fluid it is possible to explain the accelerated expansion of the Universe, we split the dark sector into two fluids, dark matter and dark energy, to satisfy the near equilibrium condition required by thermodynamics, i.e., the viscous pressure $\Pi$ must satisfy the condition $|\Pi/p|\ll 1$ at least at current epochs, where $p$ is the total equilibrium pressure of the fluids. On the other hand, by requiring $\ddot{a}>0$ for late times, the condition $-\Pi> p + \rho /3$  must be satisfied, where $\rho$ is the total energy density of the fluids. The latter is fulfilled because an acceleration phase occur as it is shown in the $q(z)$ reconstruction. Figure \ref{fig:Pipvsz} shows the evolution of $|\Pi/p|$ over the redshift range $-1<z<2.5$. It is interesting to observe that the both viscous models are far equilibrium in the past, but have a trend to go to near equilibrium at current epochs. Additionally, due to the divergence at $z=-1$, CVM and PVM ends far from an equilibrium point.

\begin{table*}
\caption{Correlation between the cosmographic parameters and the free parameters for PVM. The corresponding correlation values for CVM are in parenthesis.}
\centering
\begin{tabular}{|c|ccccccccc|}
\hline
Parameters     & $h$           & $\Omega_{m0}$ & $\lambda_0$ 	& $\lambda_1$ & $q_0$          & $j_0$         & $s_0$         & $l_0$         & $z_t$   \\ 
 \hline
$h$            &  1.00 (1.00)  &  0.01 (-0.06) &  0.06 (0.30)	& -0.04    & -0.18 (-0.35)  &  0.08 (0.30)  &  0.07 (0.33)  & -0.05 (-0.28) &  0.16 (0.35) \\ [0.7ex]
$\Omega_{m0}$  &  0.01 (-0.06) &  1.00 (1.00)  &  0.12 (0.39)	&  0.00    & -0.04 (0.00)   &  0.05 (0.37)  &  0.03 (0.22)  & -0.04 (0.70)  & -0.05 (-0.19) \\ [0.7ex]
$\lambda_0$    &  0.06 (0.30)  &  0.12 (0.39)  &  1.00 (1.00) 	&  0.96    &  0.79 (-0.92)  & -0.92 (1.00)  & -0.93 (0.98)  & -0.87 (-0.35) &  0.95 (0.83)  \\ [0.7ex]
$\lambda_1$    & -0.04      &  0.00      &  0.96 	&  1.00    &  0.93       & -0.99      & -0.99      & -0.84      &  0.85     \\
$q_0$          & -0.18 (-0.35) & -0.04 (0.00)  &  0.79 (-0.92)  &  0.93    &  1.00 (1.00)   & -0.96 (-0.93) & -0.95 (-0.98) & -0.70 (0.68)  &  0.61 (-0.98) \\ [0.7ex]
$j_0$          &  0.08 (0.30)  &  0.05 (0.37)  & -0.92 (1.00)	& -0.99    & -0.96 (-0.93)  &  1.00 (1.00)  &  1.00 (0.99)  &  0.82 (-0.37) & -0.81 (0.84)  \\ [0.7ex]
$s_0$          &  0.07 (0.33)  &  0.03 (0.22)  & -0.93 (0.98)	& -0.99    & -0.95 (-0.98)  &  1.00 (0.99)  &  1.00 (1.00)  &  0.86 (-0.51) & -0.81 (0.92)  \\ [0.7ex]
$l_0$          & -0.05 (-0.28) & -0.04 (0.70)  & -0.87 (-0.35)	& -0.84    & -0.70 (0.68)   &  0.82 (-0.37) &  0.86 (-0.51) &  1.00 (1.00)  & -0.80 (-0.80) \\  [0.7ex]
$z_t$          &  0.16 (0.35)  & -0.05 (-0.19) &  0.95 (0.83) 	&  0.85    &  0.61 (-0.98)  & -0.81 (0.84)  & -0.81 (0.92)  & -0.80 (-0.80) &  1.00 (1.00)   \\ [0.7ex]
\hline
\end{tabular}
\label{tab:corr}
\end{table*}

\section{Summary} \label{Sec:DisCon}

In this work, we study the Universe filled by one perfect fluid modelling the DE component and a non-perfect fluid describing the matter. While the DE is characterized by the traditional EoS $w=-1$, the matter follows an effective EoS given by $w_{m}=-3\xi(t) H(t)/\rho_m(t)$, where $\xi$ is the viscosity coefficient. As we mentioned, for $\xi=0$ we recover the cosmology of LCDM when the relativistic species are negligible. In this vein, we constrain the free parameters assuming that $\xi$ is constant and is a polynomial function of the redshift using the latest measurements and compilations of the Hubble parameter, type Ia supernovae and strong lensing. We find good agreement to data according to $\chi^2$-value presented in Table \ref{tab:bestfits}. Additionally, we compare statistically our models with LCDM using AIC and BIC criteria, and we find that the OHD (SNIa) sample prefers equally our viscous model than LCDM. In contrast, the joint (OHD+SNIa+SLS) data give strong evidence against the viscous models over LCDM.   
Based on the joint analysis, we present the dynamics of the four main cosmographic parameters ($q$,$j$,$s$,$l$) in the region $0<z<2.5$, in order to elucidate the differences and advantages regards to the LCDM model. In this context, the reconstruction of $j$ for both models indicates an effective dynamical DE, besides the causative of the universe acceleration in this model, is proposed as a CC behavior. Therefore $j$ parameter point us the effects of the viscosity, which characterize the model under study. 
Moreover, we find strong correlations between the viscosity parameters ($\lambda_0$ and $\lambda_1$) and those parameters that characterized the Universe dynamics ($q_0$ and $z_t$), allowing modify them according to the
viscosity property. 

Finally, we observe that both viscous models finish in a Big Rip state at $z\rightarrow -1$ as it is typical for this kind of models. In addition, we observe that the PVM presents a slightly transition to a quintessence region at late times, i.e., a viscous coefficient as function of the redshift may change the behavior of the Universe from phantom to quintessence. Further studies should be developed to address this result, which will be presented elsewhere, and also to extend the model by including a radiation component, in order to study the effects of the viscosity at the CMB epochs and the consequences of these dissipative effects at the perturbative level as in those presented in \cite{Li:2009,Velten_2011,Velten:2012uv,Velten:2014xca}. In particular, the authors in \cite{Velten:2012uv} establish the constriction $\tilde{\xi}<0.24$ (at $2\sigma$) at $z=0$ for a constant viscosity. In this sense, we estimate an upper bound of $\tilde{\xi}=9\lambda(0)<0.149$ and $\tilde{\xi}=9\lambda(0)<1.617$ at $95\%$ CL for CVM and PVM respectively. We can observe that the first is tighter than the one reported in \cite{Velten:2012uv}, while our result for PVM is less restricted than the mentioned. 
Additionally, our results for PVM are also in agreement within $3\sigma$ with the one obtained by \cite{Velten_2011} (for $\xi \sim \rho^s$), and we obtain a tighter constraint for CVM.
\vspace{6pt}


\acknowledgments{We thank the anonymous referee for thoughtful remarks and suggestions. M.A.G.-A. acknowledges support from SNI-M\'exico, CONACyT research fellow, COZCyT, Instituto Avanzado de Cosmolog\'ia (IAC) and CONICYT REDES (190147). A.H.A. thanks to the PRODEP project, Mexico for resources and financial support.}


\bibliography{main}

\begin{thebibliography}{71}%
\makeatletter
\providecommand \@ifxundefined [1]{%
 \@ifx{#1\undefined}
}%
\providecommand \@ifnum [1]{%
 \ifnum #1\expandafter \@firstoftwo
 \else \expandafter \@secondoftwo
 \fi
}%
\providecommand \@ifx [1]{%
 \ifx #1\expandafter \@firstoftwo
 \else \expandafter \@secondoftwo
 \fi
}%
\providecommand \natexlab [1]{#1}%
\providecommand \enquote  [1]{``#1''}%
\providecommand \bibnamefont  [1]{#1}%
\providecommand \bibfnamefont [1]{#1}%
\providecommand \citenamefont [1]{#1}%
\providecommand \href@noop [0]{\@secondoftwo}%
\providecommand \href [0]{\begingroup \@sanitize@url \@href}%
\providecommand \@href[1]{\@@startlink{#1}\@@href}%
\providecommand \@@href[1]{\endgroup#1\@@endlink}%
\providecommand \@sanitize@url [0]{\catcode `\\12\catcode `\$12\catcode
  `\&12\catcode `\#12\catcode `\^12\catcode `\_12\catcode `\%12\relax}%
\providecommand \@@startlink[1]{}%
\providecommand \@@endlink[0]{}%
\providecommand \url  [0]{\begingroup\@sanitize@url \@url }%
\providecommand \@url [1]{\endgroup\@href {#1}{\urlprefix }}%
\providecommand \urlprefix  [0]{URL }%
\providecommand \Eprint [0]{\href }%
\providecommand \doibase [0]{http://dx.doi.org/}%
\providecommand \selectlanguage [0]{\@gobble}%
\providecommand \bibinfo  [0]{\@secondoftwo}%
\providecommand \bibfield  [0]{\@secondoftwo}%
\providecommand \translation [1]{[#1]}%
\providecommand \BibitemOpen [0]{}%
\providecommand \bibitemStop [0]{}%
\providecommand \bibitemNoStop [0]{.\EOS\space}%
\providecommand \EOS [0]{\spacefactor3000\relax}%
\providecommand \BibitemShut  [1]{\csname bibitem#1\endcsname}%
\let\auto@bib@innerbib\@empty
\bibitem [{\citenamefont {Riess}\ \emph {et~al.}(1998)\citenamefont {Riess},
  \citenamefont {Filippenko}, \citenamefont {Challis}, \citenamefont
  {Clocchiatti}, \citenamefont {Diercks} \emph {et~al.}}]{Riess:1998}%
  \BibitemOpen
  \bibfield  {author} {\bibinfo {author} {\bibfnamefont {A.~G.}\ \bibnamefont
  {Riess}}, \bibinfo {author} {\bibfnamefont {A.~V.}\ \bibnamefont
  {Filippenko}}, \bibinfo {author} {\bibfnamefont {P.}~\bibnamefont {Challis}},
  \bibinfo {author} {\bibfnamefont {A.}~\bibnamefont {Clocchiatti}}, \bibinfo
  {author} {\bibfnamefont {A.}~\bibnamefont {Diercks}},  \emph {et~al.},\
  }\href {http://stacks.iop.org/1538-3881/116/i=3/a=1009} {\bibfield  {journal}
  {\bibinfo  {journal} {The Astronomical Journal}\ }\textbf {\bibinfo {volume}
  {116}},\ \bibinfo {pages} {1009} (\bibinfo {year} {1998})}\BibitemShut
  {NoStop}%
\bibitem [{\citenamefont {Aghanim}\ \emph {et~al.}(2018)\citenamefont {Aghanim}
  \emph {et~al.}}]{Planck:2018}%
  \BibitemOpen
  \bibfield  {author} {\bibinfo {author} {\bibfnamefont {N.}~\bibnamefont
  {Aghanim}} \emph {et~al.} (\bibinfo {collaboration} {Planck}),\ }\href@noop
  {} {\  (\bibinfo {year} {2018})},\ \Eprint {http://arxiv.org/abs/1807.06209}
  {arXiv:1807.06209 [astro-ph.CO]} \BibitemShut {NoStop}%
\bibitem [{\citenamefont {Maga\~na}\ \emph {et~al.}(2018)\citenamefont
  {Maga\~na}, \citenamefont {Amante}, \citenamefont {Garc\'ia-Aspeitia},\ and\
  \citenamefont {Motta}}]{Magana:2017nfs}%
  \BibitemOpen
  \bibfield  {author} {\bibinfo {author} {\bibfnamefont {J.}~\bibnamefont
  {Maga\~na}}, \bibinfo {author} {\bibfnamefont {M.~H.}\ \bibnamefont
  {Amante}}, \bibinfo {author} {\bibfnamefont {M.~A.}\ \bibnamefont
  {Garc\'ia-Aspeitia}}, \ and\ \bibinfo {author} {\bibfnamefont
  {V.}~\bibnamefont {Motta}},\ }\href {\doibase 10.1093/mnras/sty260}
  {\bibfield  {journal} {\bibinfo  {journal} {Mon. Not. Roy. Astron. Soc.}\
  }\textbf {\bibinfo {volume} {476}},\ \bibinfo {pages} {1036} (\bibinfo {year}
  {2018})},\ \Eprint {http://arxiv.org/abs/1706.09848} {arXiv:1706.09848
  [astro-ph.CO]} \BibitemShut {NoStop}%
\bibitem [{\citenamefont {Alam}\ \emph {et~al.}(2017)\citenamefont {Alam},
  \citenamefont {Ata}, \citenamefont {Bailey}, \citenamefont {Beutler},
  \citenamefont {Bizyaev}, \citenamefont {Blazek}, \citenamefont {Bolton},
  \citenamefont {Brownstein}, \citenamefont {Burden}, \citenamefont {Chuang},
  \citenamefont {Comparat}, \citenamefont {Cuesta}, \citenamefont {Dawson},
  \citenamefont {Eisenstein}, \citenamefont {Escoffier}, \citenamefont
  {Gil-Marín}, \citenamefont {Grieb}, \citenamefont {Hand}, \citenamefont
  {Ho}, \citenamefont {Kinemuchi}, \citenamefont {Kirkby}, \citenamefont
  {Kitaura}, \citenamefont {Malanushenko}, \citenamefont {Malanushenko},
  \citenamefont {Maraston}, \citenamefont {McBride}, \citenamefont {Nichol},
  \citenamefont {Olmstead}, \citenamefont {Oravetz}, \citenamefont
  {Padmanabhan}, \citenamefont {Palanque-Delabrouille}, \citenamefont {Pan},
  \citenamefont {Pellejero-Ibanez}, \citenamefont {Percival}, \citenamefont
  {Petitjean}, \citenamefont {Prada}, \citenamefont {Price-Whelan},
  \citenamefont {Reid}, \citenamefont {Rodríguez-Torres}, \citenamefont {Roe},
  \citenamefont {Ross}, \citenamefont {Ross}, \citenamefont {Rossi},
  \citenamefont {Rubiño-Martín}, \citenamefont {Saito}, \citenamefont
  {Salazar-Albornoz}, \citenamefont {Samushia}, \citenamefont {Sánchez},
  \citenamefont {Satpathy}, \citenamefont {Schlegel}, \citenamefont
  {Schneider}, \citenamefont {Scóccola}, \citenamefont {Seo}, \citenamefont
  {Sheldon}, \citenamefont {Simmons}, \citenamefont {Slosar}, \citenamefont
  {Strauss}, \citenamefont {Swanson}, \citenamefont {Thomas}, \citenamefont
  {Tinker}, \citenamefont {Tojeiro}, \citenamefont {Magaña}, \citenamefont
  {Vazquez}, \citenamefont {Verde}, \citenamefont {Wake}, \citenamefont {Wang},
  \citenamefont {Weinberg}, \citenamefont {White}, \citenamefont {Wood-Vasey},
  \citenamefont {Yèche}, \citenamefont {Zehavi}, \citenamefont {Zhai},\ and\
  \citenamefont {Zhao}}]{bao:2017}%
  \BibitemOpen
  \bibfield  {author} {\bibinfo {author} {\bibfnamefont {S.}~\bibnamefont
  {Alam}}, \bibinfo {author} {\bibfnamefont {M.}~\bibnamefont {Ata}}, \bibinfo
  {author} {\bibfnamefont {S.}~\bibnamefont {Bailey}}, \bibinfo {author}
  {\bibfnamefont {F.}~\bibnamefont {Beutler}}, \bibinfo {author} {\bibfnamefont
  {D.}~\bibnamefont {Bizyaev}}, \bibinfo {author} {\bibfnamefont {J.~A.}\
  \bibnamefont {Blazek}}, \bibinfo {author} {\bibfnamefont {A.~S.}\
  \bibnamefont {Bolton}}, \bibinfo {author} {\bibfnamefont {J.~R.}\
  \bibnamefont {Brownstein}}, \bibinfo {author} {\bibfnamefont
  {A.}~\bibnamefont {Burden}}, \bibinfo {author} {\bibfnamefont {C.-H.}\
  \bibnamefont {Chuang}}, \bibinfo {author} {\bibfnamefont {J.}~\bibnamefont
  {Comparat}}, \bibinfo {author} {\bibfnamefont {A.~J.}\ \bibnamefont
  {Cuesta}}, \bibinfo {author} {\bibfnamefont {K.~S.}\ \bibnamefont {Dawson}},
  \bibinfo {author} {\bibfnamefont {D.~J.}\ \bibnamefont {Eisenstein}},
  \bibinfo {author} {\bibfnamefont {S.}~\bibnamefont {Escoffier}}, \bibinfo
  {author} {\bibfnamefont {H.}~\bibnamefont {Gil-Marín}}, \bibinfo {author}
  {\bibfnamefont {J.~N.}\ \bibnamefont {Grieb}}, \bibinfo {author}
  {\bibfnamefont {N.}~\bibnamefont {Hand}}, \bibinfo {author} {\bibfnamefont
  {S.}~\bibnamefont {Ho}}, \bibinfo {author} {\bibfnamefont {K.}~\bibnamefont
  {Kinemuchi}}, \bibinfo {author} {\bibfnamefont {D.}~\bibnamefont {Kirkby}},
  \bibinfo {author} {\bibfnamefont {F.}~\bibnamefont {Kitaura}}, \bibinfo
  {author} {\bibfnamefont {E.}~\bibnamefont {Malanushenko}}, \bibinfo {author}
  {\bibfnamefont {V.}~\bibnamefont {Malanushenko}}, \bibinfo {author}
  {\bibfnamefont {C.}~\bibnamefont {Maraston}}, \bibinfo {author}
  {\bibfnamefont {C.~K.}\ \bibnamefont {McBride}}, \bibinfo {author}
  {\bibfnamefont {R.~C.}\ \bibnamefont {Nichol}}, \bibinfo {author}
  {\bibfnamefont {M.~D.}\ \bibnamefont {Olmstead}}, \bibinfo {author}
  {\bibfnamefont {D.}~\bibnamefont {Oravetz}}, \bibinfo {author} {\bibfnamefont
  {N.}~\bibnamefont {Padmanabhan}}, \bibinfo {author} {\bibfnamefont
  {N.}~\bibnamefont {Palanque-Delabrouille}}, \bibinfo {author} {\bibfnamefont
  {K.}~\bibnamefont {Pan}}, \bibinfo {author} {\bibfnamefont {M.}~\bibnamefont
  {Pellejero-Ibanez}}, \bibinfo {author} {\bibfnamefont {W.~J.}\ \bibnamefont
  {Percival}}, \bibinfo {author} {\bibfnamefont {P.}~\bibnamefont {Petitjean}},
  \bibinfo {author} {\bibfnamefont {F.}~\bibnamefont {Prada}}, \bibinfo
  {author} {\bibfnamefont {A.~M.}\ \bibnamefont {Price-Whelan}}, \bibinfo
  {author} {\bibfnamefont {B.~A.}\ \bibnamefont {Reid}}, \bibinfo {author}
  {\bibfnamefont {S.~A.}\ \bibnamefont {Rodríguez-Torres}}, \bibinfo {author}
  {\bibfnamefont {N.~A.}\ \bibnamefont {Roe}}, \bibinfo {author} {\bibfnamefont
  {A.~J.}\ \bibnamefont {Ross}}, \bibinfo {author} {\bibfnamefont {N.~P.}\
  \bibnamefont {Ross}}, \bibinfo {author} {\bibfnamefont {G.}~\bibnamefont
  {Rossi}}, \bibinfo {author} {\bibfnamefont {J.~A.}\ \bibnamefont
  {Rubiño-Martín}}, \bibinfo {author} {\bibfnamefont {S.}~\bibnamefont
  {Saito}}, \bibinfo {author} {\bibfnamefont {S.}~\bibnamefont
  {Salazar-Albornoz}}, \bibinfo {author} {\bibfnamefont {L.}~\bibnamefont
  {Samushia}}, \bibinfo {author} {\bibfnamefont {A.~G.}\ \bibnamefont
  {Sánchez}}, \bibinfo {author} {\bibfnamefont {S.}~\bibnamefont {Satpathy}},
  \bibinfo {author} {\bibfnamefont {D.~J.}\ \bibnamefont {Schlegel}}, \bibinfo
  {author} {\bibfnamefont {D.~P.}\ \bibnamefont {Schneider}}, \bibinfo {author}
  {\bibfnamefont {C.~G.}\ \bibnamefont {Scóccola}}, \bibinfo {author}
  {\bibfnamefont {H.-J.}\ \bibnamefont {Seo}}, \bibinfo {author} {\bibfnamefont
  {E.~S.}\ \bibnamefont {Sheldon}}, \bibinfo {author} {\bibfnamefont
  {A.}~\bibnamefont {Simmons}}, \bibinfo {author} {\bibfnamefont
  {A.}~\bibnamefont {Slosar}}, \bibinfo {author} {\bibfnamefont {M.~A.}\
  \bibnamefont {Strauss}}, \bibinfo {author} {\bibfnamefont {M.~E.~C.}\
  \bibnamefont {Swanson}}, \bibinfo {author} {\bibfnamefont {D.}~\bibnamefont
  {Thomas}}, \bibinfo {author} {\bibfnamefont {J.~L.}\ \bibnamefont {Tinker}},
  \bibinfo {author} {\bibfnamefont {R.}~\bibnamefont {Tojeiro}}, \bibinfo
  {author} {\bibfnamefont {M.~V.}\ \bibnamefont {Magaña}}, \bibinfo {author}
  {\bibfnamefont {J.~A.}\ \bibnamefont {Vazquez}}, \bibinfo {author}
  {\bibfnamefont {L.}~\bibnamefont {Verde}}, \bibinfo {author} {\bibfnamefont
  {D.~A.}\ \bibnamefont {Wake}}, \bibinfo {author} {\bibfnamefont
  {Y.}~\bibnamefont {Wang}}, \bibinfo {author} {\bibfnamefont {D.~H.}\
  \bibnamefont {Weinberg}}, \bibinfo {author} {\bibfnamefont {M.}~\bibnamefont
  {White}}, \bibinfo {author} {\bibfnamefont {W.~M.}\ \bibnamefont
  {Wood-Vasey}}, \bibinfo {author} {\bibfnamefont {C.}~\bibnamefont {Yèche}},
  \bibinfo {author} {\bibfnamefont {I.}~\bibnamefont {Zehavi}}, \bibinfo
  {author} {\bibfnamefont {Z.}~\bibnamefont {Zhai}}, \ and\ \bibinfo {author}
  {\bibfnamefont {G.-B.}\ \bibnamefont {Zhao}},\ }\href {\doibase
  10.1093/mnras/stx721} {\bibfield  {journal} {\bibinfo  {journal} {Monthly
  Notices of the Royal Astronomical Society}\ }\textbf {\bibinfo {volume}
  {470}},\ \bibinfo {pages} {2617} (\bibinfo {year} {2017})},\ \Eprint
  {http://arxiv.org/abs/http://oup.prod.sis.lan/mnras/article-pdf/470/3/2617/18315003/stx721.pdf}
  {http://oup.prod.sis.lan/mnras/article-pdf/470/3/2617/18315003/stx721.pdf}
  \BibitemShut {NoStop}%
\bibitem [{\citenamefont {Amante}\ \emph {et~al.}(2019)\citenamefont {Amante},
  \citenamefont {Maga\~na}, \citenamefont {Motta}, \citenamefont
  {Garc\'ia-Aspeitia},\ and\ \citenamefont {Verdugo}}]{Amante:2019xao}%
  \BibitemOpen
  \bibfield  {author} {\bibinfo {author} {\bibfnamefont {M.~H.}\ \bibnamefont
  {Amante}}, \bibinfo {author} {\bibfnamefont {J.}~\bibnamefont {Maga\~na}},
  \bibinfo {author} {\bibfnamefont {V.}~\bibnamefont {Motta}}, \bibinfo
  {author} {\bibfnamefont {M.~A.}\ \bibnamefont {Garc\'ia-Aspeitia}}, \ and\
  \bibinfo {author} {\bibfnamefont {T.}~\bibnamefont {Verdugo}},\ }\href@noop
  {} {\  (\bibinfo {year} {2019})},\ \Eprint {http://arxiv.org/abs/1906.04107}
  {arXiv:1906.04107 [astro-ph.CO]} \BibitemShut {NoStop}%
\bibitem [{\citenamefont {Klypin}\ \emph {et~al.}(1999)\citenamefont {Klypin},
  \citenamefont {Kravtsov}, \citenamefont {Valenzuela},\ and\ \citenamefont
  {Prada}}]{Klypin:1999uc}%
  \BibitemOpen
  \bibfield  {author} {\bibinfo {author} {\bibfnamefont {A.~A.}\ \bibnamefont
  {Klypin}}, \bibinfo {author} {\bibfnamefont {A.~V.}\ \bibnamefont
  {Kravtsov}}, \bibinfo {author} {\bibfnamefont {O.}~\bibnamefont
  {Valenzuela}}, \ and\ \bibinfo {author} {\bibfnamefont {F.}~\bibnamefont
  {Prada}},\ }\href {\doibase 10.1086/307643} {\bibfield  {journal} {\bibinfo
  {journal} {Astrophys. J.}\ }\textbf {\bibinfo {volume} {522}},\ \bibinfo
  {pages} {82} (\bibinfo {year} {1999})},\ \Eprint
  {http://arxiv.org/abs/astro-ph/9901240} {arXiv:astro-ph/9901240} \BibitemShut
  {NoStop}%
\bibitem [{\citenamefont {Moore}\ \emph {et~al.}(1999)\citenamefont {Moore},
  \citenamefont {Ghigna}, \citenamefont {Governato}, \citenamefont {Lake},
  \citenamefont {Quinn}, \citenamefont {Stadel},\ and\ \citenamefont
  {Tozzi}}]{Moore:1999nt}%
  \BibitemOpen
  \bibfield  {author} {\bibinfo {author} {\bibfnamefont {B.}~\bibnamefont
  {Moore}}, \bibinfo {author} {\bibfnamefont {S.}~\bibnamefont {Ghigna}},
  \bibinfo {author} {\bibfnamefont {F.}~\bibnamefont {Governato}}, \bibinfo
  {author} {\bibfnamefont {G.}~\bibnamefont {Lake}}, \bibinfo {author}
  {\bibfnamefont {T.~R.}\ \bibnamefont {Quinn}}, \bibinfo {author}
  {\bibfnamefont {J.}~\bibnamefont {Stadel}}, \ and\ \bibinfo {author}
  {\bibfnamefont {P.}~\bibnamefont {Tozzi}},\ }\href {\doibase 10.1086/312287}
  {\bibfield  {journal} {\bibinfo  {journal} {Astrophys. J.}\ }\textbf
  {\bibinfo {volume} {524}},\ \bibinfo {pages} {L19} (\bibinfo {year}
  {1999})},\ \Eprint {http://arxiv.org/abs/astro-ph/9907411}
  {arXiv:astro-ph/9907411} \BibitemShut {NoStop}%
\bibitem [{\citenamefont {{Navarro}}\ \emph {et~al.}(1996)\citenamefont
  {{Navarro}}, \citenamefont {{Frenk}},\ and\ \citenamefont
  {{White}}}]{NFW:1996}%
  \BibitemOpen
  \bibfield  {author} {\bibinfo {author} {\bibfnamefont {J.~F.}\ \bibnamefont
  {{Navarro}}}, \bibinfo {author} {\bibfnamefont {C.~S.}\ \bibnamefont
  {{Frenk}}}, \ and\ \bibinfo {author} {\bibfnamefont {S.~D.~M.}\ \bibnamefont
  {{White}}},\ }\href {\doibase 10.1086/177173} {\bibfield  {journal} {\bibinfo
   {journal} {\apj}\ }\textbf {\bibinfo {volume} {462}},\ \bibinfo {pages}
  {563} (\bibinfo {year} {1996})},\ \Eprint
  {http://arxiv.org/abs/astro-ph/9508025} {arXiv:astro-ph/9508025 [astro-ph]}
  \BibitemShut {NoStop}%
\bibitem [{\citenamefont {Weinberg}(1989)}]{RevModPhys.61.1}%
  \BibitemOpen
  \bibfield  {author} {\bibinfo {author} {\bibfnamefont {S.}~\bibnamefont
  {Weinberg}},\ }\href {\doibase 10.1103/RevModPhys.61.1} {\bibfield  {journal}
  {\bibinfo  {journal} {Rev. Mod. Phys.}\ }\textbf {\bibinfo {volume} {61}},\
  \bibinfo {pages} {1} (\bibinfo {year} {1989})}\BibitemShut {NoStop}%
\bibitem [{\citenamefont {Astashenok}\ and\ \citenamefont {del
  Popolo}(2012)}]{Astashenok_2012}%
  \BibitemOpen
  \bibfield  {author} {\bibinfo {author} {\bibfnamefont {A.~V.}\ \bibnamefont
  {Astashenok}}\ and\ \bibinfo {author} {\bibfnamefont {A.}~\bibnamefont {del
  Popolo}},\ }\href {\doibase 10.1088/0264-9381/29/8/085014} {\bibfield
  {journal} {\bibinfo  {journal} {Classical and Quantum Gravity}\ }\textbf
  {\bibinfo {volume} {29}},\ \bibinfo {pages} {085014} (\bibinfo {year}
  {2012})}\BibitemShut {NoStop}%
\bibitem [{\citenamefont {Martin}(2012)}]{martin2012everything}%
  \BibitemOpen
  \bibfield  {author} {\bibinfo {author} {\bibfnamefont {J.}~\bibnamefont
  {Martin}},\ }\href@noop {} {\bibfield  {journal} {\bibinfo  {journal}
  {Comptes Rendus Physique}\ }\textbf {\bibinfo {volume} {13}},\ \bibinfo
  {pages} {566} (\bibinfo {year} {2012})}\BibitemShut {NoStop}%
\bibitem [{\citenamefont {Del~Popolo}\ and\ \citenamefont
  {Le~Delliou}(2017)}]{del2017small}%
  \BibitemOpen
  \bibfield  {author} {\bibinfo {author} {\bibfnamefont {A.}~\bibnamefont
  {Del~Popolo}}\ and\ \bibinfo {author} {\bibfnamefont {M.}~\bibnamefont
  {Le~Delliou}},\ }\href@noop {} {\bibfield  {journal} {\bibinfo  {journal}
  {Galaxies}\ }\textbf {\bibinfo {volume} {5}},\ \bibinfo {pages} {17}
  (\bibinfo {year} {2017})}\BibitemShut {NoStop}%
\bibitem [{\citenamefont {Garcia-Aspeitia}\ \emph {et~al.}(2018)\citenamefont
  {Garcia-Aspeitia}, \citenamefont {Hernandez-Almada}, \citenamefont {Magaña},
  \citenamefont {Amante}, \citenamefont {Motta},\ and\ \citenamefont
  {Martínez-Robles}}]{Garcia-Aspeitia:2018fvw}%
  \BibitemOpen
  \bibfield  {author} {\bibinfo {author} {\bibfnamefont {M.~A.}\ \bibnamefont
  {Garcia-Aspeitia}}, \bibinfo {author} {\bibfnamefont {A.}~\bibnamefont
  {Hernandez-Almada}}, \bibinfo {author} {\bibfnamefont {J.}~\bibnamefont
  {Magaña}}, \bibinfo {author} {\bibfnamefont {M.~H.}\ \bibnamefont {Amante}},
  \bibinfo {author} {\bibfnamefont {V.}~\bibnamefont {Motta}}, \ and\ \bibinfo
  {author} {\bibfnamefont {C.}~\bibnamefont {Martínez-Robles}},\ }\href
  {\doibase 10.1103/PhysRevD.97.101301} {\bibfield  {journal} {\bibinfo
  {journal} {Phys. Rev. D}\ }\textbf {\bibinfo {volume} {97}},\ \bibinfo
  {pages} {101301} (\bibinfo {year} {2018})},\ \Eprint
  {http://arxiv.org/abs/1804.05085} {arXiv:1804.05085 [gr-qc]} \BibitemShut
  {NoStop}%
\bibitem [{\citenamefont {Hernandez-Almada}\ \emph {et~al.}(2019)\citenamefont
  {Hernandez-Almada}, \citenamefont {Magana}, \citenamefont {Garcia-Aspeitia},\
  and\ \citenamefont {Motta}}]{Hernandez-Almada:2018osh}%
  \BibitemOpen
  \bibfield  {author} {\bibinfo {author} {\bibfnamefont {A.}~\bibnamefont
  {Hernandez-Almada}}, \bibinfo {author} {\bibfnamefont {J.}~\bibnamefont
  {Magana}}, \bibinfo {author} {\bibfnamefont {M.~A.}\ \bibnamefont
  {Garcia-Aspeitia}}, \ and\ \bibinfo {author} {\bibfnamefont {V.}~\bibnamefont
  {Motta}},\ }\href {\doibase 10.1140/epjc/s10052-018-6521-6} {\bibfield
  {journal} {\bibinfo  {journal} {Eur. Phys. J. C}\ }\textbf {\bibinfo {volume}
  {79}},\ \bibinfo {pages} {12} (\bibinfo {year} {2019})},\ \Eprint
  {http://arxiv.org/abs/1805.07895} {arXiv:1805.07895 [astro-ph.CO]}
  \BibitemShut {NoStop}%
\bibitem [{\citenamefont {García-Aspeitia}\ \emph
  {et~al.}(2019{\natexlab{a}})\citenamefont {García-Aspeitia}, \citenamefont
  {Hernández-Almada}, \citenamefont {Magaña},\ and\ \citenamefont
  {Motta}}]{Garcia-Aspeitia:2019yod}%
  \BibitemOpen
  \bibfield  {author} {\bibinfo {author} {\bibfnamefont {M.~A.}\ \bibnamefont
  {García-Aspeitia}}, \bibinfo {author} {\bibfnamefont {A.}~\bibnamefont
  {Hernández-Almada}}, \bibinfo {author} {\bibfnamefont {J.}~\bibnamefont
  {Magaña}}, \ and\ \bibinfo {author} {\bibfnamefont {V.}~\bibnamefont
  {Motta}},\ }\href@noop {} {\  (\bibinfo {year} {2019}{\natexlab{a}})},\
  \Eprint {http://arxiv.org/abs/1912.07500} {arXiv:1912.07500 [astro-ph.CO]}
  \BibitemShut {NoStop}%
\bibitem [{\citenamefont {García-Aspeitia}\ \emph
  {et~al.}(2019{\natexlab{b}})\citenamefont {García-Aspeitia}, \citenamefont
  {Martínez-Robles}, \citenamefont {Hernández-Almada}, \citenamefont
  {Magaña},\ and\ \citenamefont {Motta}}]{Garcia-Aspeitia:2019yni}%
  \BibitemOpen
  \bibfield  {author} {\bibinfo {author} {\bibfnamefont {M.~A.}\ \bibnamefont
  {García-Aspeitia}}, \bibinfo {author} {\bibfnamefont {C.}~\bibnamefont
  {Martínez-Robles}}, \bibinfo {author} {\bibfnamefont {A.}~\bibnamefont
  {Hernández-Almada}}, \bibinfo {author} {\bibfnamefont {J.}~\bibnamefont
  {Magaña}}, \ and\ \bibinfo {author} {\bibfnamefont {V.}~\bibnamefont
  {Motta}},\ }\href {\doibase 10.1103/PhysRevD.99.123525} {\bibfield  {journal}
  {\bibinfo  {journal} {Phys. Rev. D}\ }\textbf {\bibinfo {volume} {99}},\
  \bibinfo {pages} {123525} (\bibinfo {year} {2019}{\natexlab{b}})},\ \Eprint
  {http://arxiv.org/abs/1903.06344} {arXiv:1903.06344 [gr-qc]} \BibitemShut
  {NoStop}%
\bibitem [{\citenamefont {Copeland}\ \emph {et~al.}(2006)\citenamefont
  {Copeland}, \citenamefont {Sami},\ and\ \citenamefont
  {Tsujikawa}}]{Copeland:2006wr}%
  \BibitemOpen
  \bibfield  {author} {\bibinfo {author} {\bibfnamefont {E.~J.}\ \bibnamefont
  {Copeland}}, \bibinfo {author} {\bibfnamefont {M.}~\bibnamefont {Sami}}, \
  and\ \bibinfo {author} {\bibfnamefont {S.}~\bibnamefont {Tsujikawa}},\ }\href
  {\doibase 10.1142/S021827180600942X} {\bibfield  {journal} {\bibinfo
  {journal} {Int. J. Mod. Phys. D}\ }\textbf {\bibinfo {volume} {15}},\
  \bibinfo {pages} {1753} (\bibinfo {year} {2006})},\ \Eprint
  {http://arxiv.org/abs/hep-th/0603057} {arXiv:hep-th/0603057} \BibitemShut
  {NoStop}%
\bibitem [{\citenamefont {Matos}\ and\ \citenamefont
  {Urena-Lopez}(2001)}]{Matos:2000ss}%
  \BibitemOpen
  \bibfield  {author} {\bibinfo {author} {\bibfnamefont {T.}~\bibnamefont
  {Matos}}\ and\ \bibinfo {author} {\bibfnamefont {L.~A.}\ \bibnamefont
  {Urena-Lopez}},\ }\href {\doibase 10.1103/PhysRevD.63.063506} {\bibfield
  {journal} {\bibinfo  {journal} {Phys. Rev. D}\ }\textbf {\bibinfo {volume}
  {63}},\ \bibinfo {pages} {063506} (\bibinfo {year} {2001})},\ \Eprint
  {http://arxiv.org/abs/astro-ph/0006024} {arXiv:astro-ph/0006024} \BibitemShut
  {NoStop}%
\bibitem [{\citenamefont {Matos}\ and\ \citenamefont
  {Urena-Lopez}(2000)}]{Matos:2000ng}%
  \BibitemOpen
  \bibfield  {author} {\bibinfo {author} {\bibfnamefont {T.}~\bibnamefont
  {Matos}}\ and\ \bibinfo {author} {\bibfnamefont {L.}~\bibnamefont
  {Urena-Lopez}},\ }\href {\doibase 10.1088/0264-9381/17/13/101} {\bibfield
  {journal} {\bibinfo  {journal} {Class. Quant. Grav.}\ }\textbf {\bibinfo
  {volume} {17}},\ \bibinfo {pages} {L75} (\bibinfo {year} {2000})},\ \Eprint
  {http://arxiv.org/abs/astro-ph/0004332} {arXiv:astro-ph/0004332} \BibitemShut
  {NoStop}%
\bibitem [{\citenamefont {Matos}\ \emph {et~al.}(2000)\citenamefont {Matos},
  \citenamefont {Guzman},\ and\ \citenamefont {Nunez}}]{Matos:2000ki}%
  \BibitemOpen
  \bibfield  {author} {\bibinfo {author} {\bibfnamefont {T.}~\bibnamefont
  {Matos}}, \bibinfo {author} {\bibfnamefont {F.~S.}\ \bibnamefont {Guzman}}, \
  and\ \bibinfo {author} {\bibfnamefont {D.}~\bibnamefont {Nunez}},\ }\href
  {\doibase 10.1103/PhysRevD.62.061301} {\bibfield  {journal} {\bibinfo
  {journal} {Phys. Rev. D}\ }\textbf {\bibinfo {volume} {62}},\ \bibinfo
  {pages} {061301} (\bibinfo {year} {2000})},\ \Eprint
  {http://arxiv.org/abs/astro-ph/0003398} {arXiv:astro-ph/0003398} \BibitemShut
  {NoStop}%
\bibitem [{\citenamefont {Urena-Lopez}\ and\ \citenamefont
  {Matos}(2000)}]{UrenaLopez:2000aj}%
  \BibitemOpen
  \bibfield  {author} {\bibinfo {author} {\bibfnamefont {L.}~\bibnamefont
  {Urena-Lopez}}\ and\ \bibinfo {author} {\bibfnamefont {T.}~\bibnamefont
  {Matos}},\ }\href {\doibase 10.1103/PhysRevD.62.081302} {\bibfield  {journal}
  {\bibinfo  {journal} {Phys. Rev. D}\ }\textbf {\bibinfo {volume} {62}},\
  \bibinfo {pages} {081302} (\bibinfo {year} {2000})},\ \Eprint
  {http://arxiv.org/abs/astro-ph/0003364} {arXiv:astro-ph/0003364} \BibitemShut
  {NoStop}%
\bibitem [{\citenamefont {Peccei}(2008)}]{Peccei:2006as}%
  \BibitemOpen
  \bibfield  {author} {\bibinfo {author} {\bibfnamefont {R.}~\bibnamefont
  {Peccei}},\ }\href {\doibase 10.1007/978-3-540-73518-2\_1} {\bibfield
  {journal} {\bibinfo  {journal} {Lect. Notes Phys.}\ }\textbf {\bibinfo
  {volume} {741}},\ \bibinfo {pages} {3} (\bibinfo {year} {2008})},\ \Eprint
  {http://arxiv.org/abs/hep-ph/0607268} {arXiv:hep-ph/0607268} \BibitemShut
  {NoStop}%
\bibitem [{\citenamefont {Berenji}\ \emph {et~al.}(2016)\citenamefont
  {Berenji}, \citenamefont {Gaskins},\ and\ \citenamefont
  {Meyer}}]{Berenji:2016jji}%
  \BibitemOpen
  \bibfield  {author} {\bibinfo {author} {\bibfnamefont {B.}~\bibnamefont
  {Berenji}}, \bibinfo {author} {\bibfnamefont {J.}~\bibnamefont {Gaskins}}, \
  and\ \bibinfo {author} {\bibfnamefont {M.}~\bibnamefont {Meyer}},\ }\href
  {\doibase 10.1103/PhysRevD.93.045019} {\bibfield  {journal} {\bibinfo
  {journal} {Phys. Rev. D}\ }\textbf {\bibinfo {volume} {93}},\ \bibinfo
  {pages} {045019} (\bibinfo {year} {2016})},\ \Eprint
  {http://arxiv.org/abs/1602.00091} {arXiv:1602.00091 [astro-ph.HE]}
  \BibitemShut {NoStop}%
\bibitem [{\citenamefont {Avelino}\ \emph {et~al.}(2013)\citenamefont
  {Avelino}, \citenamefont {Leyva},\ and\ \citenamefont {Ure\~na
  L\'opez}}]{Avelino:2013}%
  \BibitemOpen
  \bibfield  {author} {\bibinfo {author} {\bibfnamefont {A.}~\bibnamefont
  {Avelino}}, \bibinfo {author} {\bibfnamefont {Y.}~\bibnamefont {Leyva}}, \
  and\ \bibinfo {author} {\bibfnamefont {L.~A.}\ \bibnamefont {Ure\~na
  L\'opez}},\ }\href {\doibase 10.1103/PhysRevD.88.123004} {\bibfield
  {journal} {\bibinfo  {journal} {Phys. Rev. D}\ }\textbf {\bibinfo {volume}
  {88}},\ \bibinfo {pages} {123004} (\bibinfo {year} {2013})}\BibitemShut
  {NoStop}%
\bibitem [{\citenamefont {Atreya}\ \emph {et~al.}(2018)\citenamefont {Atreya},
  \citenamefont {Bhatt},\ and\ \citenamefont {Mishra}}]{Atreya_2018}%
  \BibitemOpen
  \bibfield  {author} {\bibinfo {author} {\bibfnamefont {A.}~\bibnamefont
  {Atreya}}, \bibinfo {author} {\bibfnamefont {J.~R.}\ \bibnamefont {Bhatt}}, \
  and\ \bibinfo {author} {\bibfnamefont {A.}~\bibnamefont {Mishra}},\ }\href
  {\doibase 10.1088/1475-7516/2018/02/024} {\bibfield  {journal} {\bibinfo
  {journal} {Journal of Cosmology and Astroparticle Physics}\ }\textbf
  {\bibinfo {volume} {2018}},\ \bibinfo {pages} {024} (\bibinfo {year}
  {2018})}\BibitemShut {NoStop}%
\bibitem [{\citenamefont {Di~Valentino}\ \emph {et~al.}(2019)\citenamefont
  {Di~Valentino}, \citenamefont {Melchiorri}, \citenamefont {Mena},\ and\
  \citenamefont {Vagnozzi}}]{DiValentino:2019ffd}%
  \BibitemOpen
  \bibfield  {author} {\bibinfo {author} {\bibfnamefont {E.}~\bibnamefont
  {Di~Valentino}}, \bibinfo {author} {\bibfnamefont {A.}~\bibnamefont
  {Melchiorri}}, \bibinfo {author} {\bibfnamefont {O.}~\bibnamefont {Mena}}, \
  and\ \bibinfo {author} {\bibfnamefont {S.}~\bibnamefont {Vagnozzi}},\
  }\href@noop {} {\  (\bibinfo {year} {2019})},\ \Eprint
  {http://arxiv.org/abs/1908.04281} {arXiv:1908.04281 [astro-ph.CO]}
  \BibitemShut {NoStop}%
\bibitem [{\citenamefont {Bowman}\ \emph {et~al.}(2018)\citenamefont {Bowman},
  \citenamefont {Rogers}, \citenamefont {Monsalve}, \citenamefont {Mozdzen},\
  and\ \citenamefont {Mahesh}}]{Bowman:2018yin}%
  \BibitemOpen
  \bibfield  {author} {\bibinfo {author} {\bibfnamefont {J.~D.}\ \bibnamefont
  {Bowman}}, \bibinfo {author} {\bibfnamefont {A.~E.~E.}\ \bibnamefont
  {Rogers}}, \bibinfo {author} {\bibfnamefont {R.~A.}\ \bibnamefont
  {Monsalve}}, \bibinfo {author} {\bibfnamefont {T.~J.}\ \bibnamefont
  {Mozdzen}}, \ and\ \bibinfo {author} {\bibfnamefont {N.}~\bibnamefont
  {Mahesh}},\ }\href {\doibase 10.1038/nature25792} {\bibfield  {journal}
  {\bibinfo  {journal} {Nature}\ }\textbf {\bibinfo {volume} {555}},\ \bibinfo
  {pages} {67} (\bibinfo {year} {2018})},\ \Eprint
  {http://arxiv.org/abs/1810.05912} {arXiv:1810.05912 [astro-ph.CO]}
  \BibitemShut {NoStop}%
\bibitem [{\citenamefont {Eckart}(1940)}]{Eckart:1940}%
  \BibitemOpen
  \bibfield  {author} {\bibinfo {author} {\bibfnamefont {C.}~\bibnamefont
  {Eckart}},\ }\href {\doibase 10.1103/PhysRev.58.919} {\bibfield  {journal}
  {\bibinfo  {journal} {Phys. Rev.}\ }\textbf {\bibinfo {volume} {58}},\
  \bibinfo {pages} {919} (\bibinfo {year} {1940})}\BibitemShut {NoStop}%
\bibitem [{\citenamefont {Israel}\ and\ \citenamefont
  {Stewart}(1979)}]{Israel1979}%
  \BibitemOpen
  \bibfield  {author} {\bibinfo {author} {\bibfnamefont {W.}~\bibnamefont
  {Israel}}\ and\ \bibinfo {author} {\bibfnamefont {J.}~\bibnamefont
  {Stewart}},\ }\href {\doibase https://doi.org/10.1016/0003-4916(79)90130-1}
  {\bibfield  {journal} {\bibinfo  {journal} {Annals of Physics}\ }\textbf
  {\bibinfo {volume} {118}},\ \bibinfo {pages} {341 } (\bibinfo {year}
  {1979})}\BibitemShut {NoStop}%
\bibitem [{\citenamefont {Murphy}(1973)}]{Murphy:1973}%
  \BibitemOpen
  \bibfield  {author} {\bibinfo {author} {\bibfnamefont {G.~L.}\ \bibnamefont
  {Murphy}},\ }\href {\doibase 10.1103/PhysRevD.8.4231} {\bibfield  {journal}
  {\bibinfo  {journal} {Phys. Rev. D}\ }\textbf {\bibinfo {volume} {8}},\
  \bibinfo {pages} {4231} (\bibinfo {year} {1973})}\BibitemShut {NoStop}%
\bibitem [{\citenamefont {Padmanabhan}\ and\ \citenamefont
  {Chitre}(1987)}]{Padmanabhan:1987}%
  \BibitemOpen
  \bibfield  {author} {\bibinfo {author} {\bibfnamefont {T.}~\bibnamefont
  {Padmanabhan}}\ and\ \bibinfo {author} {\bibfnamefont {S.}~\bibnamefont
  {Chitre}},\ }\href {\doibase https://doi.org/10.1016/0375-9601(87)90104-6}
  {\bibfield  {journal} {\bibinfo  {journal} {Physics Letters A}\ }\textbf
  {\bibinfo {volume} {120}},\ \bibinfo {pages} {433 } (\bibinfo {year}
  {1987})}\BibitemShut {NoStop}%
\bibitem [{\citenamefont {Brevik}\ and\ \citenamefont
  {Gorbunova}(2005)}]{Brevik2005}%
  \BibitemOpen
  \bibfield  {author} {\bibinfo {author} {\bibfnamefont {I.}~\bibnamefont
  {Brevik}}\ and\ \bibinfo {author} {\bibfnamefont {O.}~\bibnamefont
  {Gorbunova}},\ }\href {\doibase 10.1007/s10714-005-0178-9} {\bibfield
  {journal} {\bibinfo  {journal} {General Relativity and Gravitation}\ }\textbf
  {\bibinfo {volume} {37}},\ \bibinfo {pages} {2039} (\bibinfo {year}
  {2005})}\BibitemShut {NoStop}%
\bibitem [{\citenamefont {Normann}\ and\ \citenamefont
  {Brevik}(2017)}]{Normann:2017}%
  \BibitemOpen
  \bibfield  {author} {\bibinfo {author} {\bibfnamefont {B.~D.}\ \bibnamefont
  {Normann}}\ and\ \bibinfo {author} {\bibfnamefont {I.}~\bibnamefont
  {Brevik}},\ }\href {\doibase 10.1142/S0217732317500262} {\bibfield  {journal}
  {\bibinfo  {journal} {Modern Physics Letters A}\ }\textbf {\bibinfo {volume}
  {32}},\ \bibinfo {pages} {1750026} (\bibinfo {year} {2017})},\ \Eprint
  {http://arxiv.org/abs/https://doi.org/10.1142/S0217732317500262}
  {https://doi.org/10.1142/S0217732317500262} \BibitemShut {NoStop}%
\bibitem [{\citenamefont {Xin-He}\ and\ \citenamefont
  {Xu}(2009)}]{Xin-He:2009}%
  \BibitemOpen
  \bibfield  {author} {\bibinfo {author} {\bibfnamefont {M.}~\bibnamefont
  {Xin-He}}\ and\ \bibinfo {author} {\bibfnamefont {D.}~\bibnamefont {Xu}},\
  }\href {http://stacks.iop.org/0253-6102/52/i=2/a=36} {\bibfield  {journal}
  {\bibinfo  {journal} {Communications in Theoretical Physics}\ }\textbf
  {\bibinfo {volume} {52}},\ \bibinfo {pages} {377} (\bibinfo {year}
  {2009})}\BibitemShut {NoStop}%
\bibitem [{\citenamefont {Avelino}\ and\ \citenamefont
  {Nucamendi}(2010)}]{Avelino:2010}%
  \BibitemOpen
  \bibfield  {author} {\bibinfo {author} {\bibfnamefont {A.}~\bibnamefont
  {Avelino}}\ and\ \bibinfo {author} {\bibfnamefont {U.}~\bibnamefont
  {Nucamendi}},\ }\href {\doibase 10.1088/1475-7516/2010/08/009} {\bibfield
  {journal} {\bibinfo  {journal} {Journal of Cosmology and Astroparticle
  Physics}\ }\textbf {\bibinfo {volume} {2010}},\ \bibinfo {pages} {009}
  (\bibinfo {year} {2010})}\BibitemShut {NoStop}%
\bibitem [{\citenamefont
  {Hern{\'a}ndez-Almada}(2019{\natexlab{a}})}]{Almada:2019}%
  \BibitemOpen
  \bibfield  {author} {\bibinfo {author} {\bibfnamefont {A.}~\bibnamefont
  {Hern{\'a}ndez-Almada}},\ }\href {\doibase 10.1140/epjc/s10052-019-7264-8}
  {\bibfield  {journal} {\bibinfo  {journal} {The European Physical Journal C}\
  }\textbf {\bibinfo {volume} {79}},\ \bibinfo {pages} {751} (\bibinfo {year}
  {2019}{\natexlab{a}})}\BibitemShut {NoStop}%
\bibitem [{\citenamefont {Folomeev}\ and\ \citenamefont
  {Gurovich}(2008)}]{FOLOMEEV200875}%
  \BibitemOpen
  \bibfield  {author} {\bibinfo {author} {\bibfnamefont {V.}~\bibnamefont
  {Folomeev}}\ and\ \bibinfo {author} {\bibfnamefont {V.}~\bibnamefont
  {Gurovich}},\ }\href {\doibase
  https://doi.org/10.1016/j.physletb.2008.01.068} {\bibfield  {journal}
  {\bibinfo  {journal} {Physics Letters B}\ }\textbf {\bibinfo {volume}
  {661}},\ \bibinfo {pages} {75 } (\bibinfo {year} {2008})}\BibitemShut
  {NoStop}%
\bibitem [{\citenamefont {Hern\'andez-Almada}\ \emph
  {et~al.}(2020)\citenamefont {Hern\'andez-Almada}, \citenamefont
  {Garc\'{\i}a-Aspeitia}, \citenamefont {Maga\~na},\ and\ \citenamefont
  {Motta}}]{Almada:2020}%
  \BibitemOpen
  \bibfield  {author} {\bibinfo {author} {\bibfnamefont {A.}~\bibnamefont
  {Hern\'andez-Almada}}, \bibinfo {author} {\bibfnamefont {M.~A.}\ \bibnamefont
  {Garc\'{\i}a-Aspeitia}}, \bibinfo {author} {\bibfnamefont {J.}~\bibnamefont
  {Maga\~na}}, \ and\ \bibinfo {author} {\bibfnamefont {V.}~\bibnamefont
  {Motta}},\ }\href {\doibase 10.1103/PhysRevD.101.063516} {\bibfield
  {journal} {\bibinfo  {journal} {Phys. Rev. D}\ }\textbf {\bibinfo {volume}
  {101}},\ \bibinfo {pages} {063516} (\bibinfo {year} {2020})}\BibitemShut
  {NoStop}%
\bibitem [{\citenamefont {CORNEJO-P\'EREZ}\ and\ \citenamefont
  {BELINCH\'ON}(2013)}]{Cornejo:2013}%
  \BibitemOpen
  \bibfield  {author} {\bibinfo {author} {\bibfnamefont {O.}~\bibnamefont
  {CORNEJO-P\'EREZ}}\ and\ \bibinfo {author} {\bibfnamefont {J.~A.}\
  \bibnamefont {BELINCH\'ON}},\ }\href {\doibase 10.1142/S0218271813500314}
  {\bibfield  {journal} {\bibinfo  {journal} {International Journal of Modern
  Physics D}\ }\textbf {\bibinfo {volume} {22}},\ \bibinfo {pages} {1350031}
  (\bibinfo {year} {2013})},\ \Eprint
  {http://arxiv.org/abs/https://doi.org/10.1142/S0218271813500314}
  {https://doi.org/10.1142/S0218271813500314} \BibitemShut {NoStop}%
\bibitem [{\citenamefont {Cruz}\ \emph
  {et~al.}(2017{\natexlab{a}})\citenamefont {Cruz}, \citenamefont {Cruz},\ and\
  \citenamefont {Lepe}}]{MCruz:2017}%
  \BibitemOpen
  \bibfield  {author} {\bibinfo {author} {\bibfnamefont {M.}~\bibnamefont
  {Cruz}}, \bibinfo {author} {\bibfnamefont {N.}~\bibnamefont {Cruz}}, \ and\
  \bibinfo {author} {\bibfnamefont {S.}~\bibnamefont {Lepe}},\ }\href {\doibase
  10.1103/PhysRevD.96.124020} {\bibfield  {journal} {\bibinfo  {journal} {Phys.
  Rev. D}\ }\textbf {\bibinfo {volume} {96}},\ \bibinfo {pages} {124020}
  (\bibinfo {year} {2017}{\natexlab{a}})}\BibitemShut {NoStop}%
\bibitem [{\citenamefont {Cruz}\ \emph
  {et~al.}(2017{\natexlab{b}})\citenamefont {Cruz}, \citenamefont {Cruz},\ and\
  \citenamefont {Lepe}}]{CRUZ2017159}%
  \BibitemOpen
  \bibfield  {author} {\bibinfo {author} {\bibfnamefont {M.}~\bibnamefont
  {Cruz}}, \bibinfo {author} {\bibfnamefont {N.}~\bibnamefont {Cruz}}, \ and\
  \bibinfo {author} {\bibfnamefont {S.}~\bibnamefont {Lepe}},\ }\href {\doibase
  https://doi.org/10.1016/j.physletb.2017.03.065} {\bibfield  {journal}
  {\bibinfo  {journal} {Physics Letters B}\ }\textbf {\bibinfo {volume}
  {769}},\ \bibinfo {pages} {159 } (\bibinfo {year}
  {2017}{\natexlab{b}})}\BibitemShut {NoStop}%
\bibitem [{\citenamefont {Norman~Cruz}(2018)}]{NCruz:2018arx}%
  \BibitemOpen
  \bibfield  {author} {\bibinfo {author} {\bibfnamefont {G.~P.}\ \bibnamefont
  {Norman~Cruz}, \bibfnamefont {Esteban~Gonz\'alez}},\ }\href@noop {} {\
  (\bibinfo {year} {2018})},\ \Eprint {http://arxiv.org/abs/1812.05009}
  {arXiv:1812.05009 [astro-ph]} \BibitemShut {NoStop}%
\bibitem [{\citenamefont {Cruz}\ \emph
  {et~al.}(2019{\natexlab{a}})\citenamefont {Cruz}, \citenamefont
  {Hern\'andez-Almada},\ and\ \citenamefont {Cornejo-P\'erez}}]{NCruz:2019}%
  \BibitemOpen
  \bibfield  {author} {\bibinfo {author} {\bibfnamefont {N.}~\bibnamefont
  {Cruz}}, \bibinfo {author} {\bibfnamefont {A.}~\bibnamefont
  {Hern\'andez-Almada}}, \ and\ \bibinfo {author} {\bibfnamefont
  {O.}~\bibnamefont {Cornejo-P\'erez}},\ }\href {\doibase
  10.1103/PhysRevD.100.083524} {\bibfield  {journal} {\bibinfo  {journal}
  {Phys. Rev. D}\ }\textbf {\bibinfo {volume} {100}},\ \bibinfo {pages}
  {083524} (\bibinfo {year} {2019}{\natexlab{a}})}\BibitemShut {NoStop}%
\bibitem [{\citenamefont {Cruz}\ \emph
  {et~al.}(2019{\natexlab{b}})\citenamefont {Cruz}, \citenamefont {Cruz},\ and\
  \citenamefont {Lepe}}]{cruz2019unified}%
  \BibitemOpen
  \bibfield  {author} {\bibinfo {author} {\bibfnamefont {M.}~\bibnamefont
  {Cruz}}, \bibinfo {author} {\bibfnamefont {N.}~\bibnamefont {Cruz}}, \ and\
  \bibinfo {author} {\bibfnamefont {S.}~\bibnamefont {Lepe}},\ }\href@noop {}
  {\enquote {\bibinfo {title} {Unified dissipative dark matter model in a non
  linear israel-stewart theory},}\ } (\bibinfo {year} {2019}{\natexlab{b}}),\
  \Eprint {http://arxiv.org/abs/1911.04539} {arXiv:1911.04539 [gr-qc]}
  \BibitemShut {NoStop}%
\bibitem [{\citenamefont {Cruz}\ \emph {et~al.}(2018)\citenamefont {Cruz},
  \citenamefont {Gonz{\'{a}}lez}, \citenamefont {Lepe},\ and\ \citenamefont
  {G{\'{o}}mez}}]{NCruz:2018}%
  \BibitemOpen
  \bibfield  {author} {\bibinfo {author} {\bibfnamefont {N.}~\bibnamefont
  {Cruz}}, \bibinfo {author} {\bibfnamefont {E.}~\bibnamefont
  {Gonz{\'{a}}lez}}, \bibinfo {author} {\bibfnamefont {S.}~\bibnamefont
  {Lepe}}, \ and\ \bibinfo {author} {\bibfnamefont {D.~S.-C.}\ \bibnamefont
  {G{\'{o}}mez}},\ }\href {\doibase 10.1088/1475-7516/2018/12/017} {\bibfield
  {journal} {\bibinfo  {journal} {Journal of Cosmology and Astroparticle
  Physics}\ }\textbf {\bibinfo {volume} {2018}},\ \bibinfo {pages} {017}
  (\bibinfo {year} {2018})}\BibitemShut {NoStop}%
\bibitem [{\citenamefont {Scolnic}\ and\ \citenamefont {{\it et.
  al.}}(2018)}]{Scolnic:2017caz}%
  \BibitemOpen
  \bibfield  {author} {\bibinfo {author} {\bibfnamefont {D.~M.}\ \bibnamefont
  {Scolnic}}\ and\ \bibinfo {author} {\bibnamefont {{\it et. al.}}},\ }\href
  {http://stacks.iop.org/0004-637X/859/i=2/a=101} {\bibfield  {journal}
  {\bibinfo  {journal} {The Astrophysical Journal}\ }\textbf {\bibinfo {volume}
  {859}},\ \bibinfo {pages} {101} (\bibinfo {year} {2018})}\BibitemShut
  {NoStop}%
\bibitem [{\citenamefont {Normann}\ and\ \citenamefont
  {Brevik}(2016)}]{Normann:2016}%
  \BibitemOpen
  \bibfield  {author} {\bibinfo {author} {\bibfnamefont {B.~D.}\ \bibnamefont
  {Normann}}\ and\ \bibinfo {author} {\bibfnamefont {I.}~\bibnamefont
  {Brevik}},\ }\href {\doibase 10.3390/e18060215} {\bibfield  {journal}
  {\bibinfo  {journal} {Entropy}\ }\textbf {\bibinfo {volume} {18}} (\bibinfo
  {year} {2016}),\ 10.3390/e18060215}\BibitemShut {NoStop}%
\bibitem [{\citenamefont
  {Hern{\'a}ndez-Almada}(2019{\natexlab{b}})}]{Hernandez-Almada:2019}%
  \BibitemOpen
  \bibfield  {author} {\bibinfo {author} {\bibfnamefont {A.}~\bibnamefont
  {Hern{\'a}ndez-Almada}},\ }\href {\doibase 10.1140/epjc/s10052-019-7264-8}
  {\bibfield  {journal} {\bibinfo  {journal} {The European Physical Journal C}\
  }\textbf {\bibinfo {volume} {79}},\ \bibinfo {pages} {751} (\bibinfo {year}
  {2019}{\natexlab{b}})}\BibitemShut {NoStop}%
\bibitem [{\citenamefont {Brevik}\ \emph {et~al.}(2011)\citenamefont {Brevik},
  \citenamefont {Elizalde}, \citenamefont {Nojiri},\ and\ \citenamefont
  {Odintsov}}]{Brevik:2011mm}%
  \BibitemOpen
  \bibfield  {author} {\bibinfo {author} {\bibfnamefont {I.}~\bibnamefont
  {Brevik}}, \bibinfo {author} {\bibfnamefont {E.}~\bibnamefont {Elizalde}},
  \bibinfo {author} {\bibfnamefont {S.}~\bibnamefont {Nojiri}}, \ and\ \bibinfo
  {author} {\bibfnamefont {S.}~\bibnamefont {Odintsov}},\ }\href {\doibase
  10.1103/PhysRevD.84.103508} {\bibfield  {journal} {\bibinfo  {journal} {Phys.
  Rev. D}\ }\textbf {\bibinfo {volume} {84}},\ \bibinfo {pages} {103508}
  (\bibinfo {year} {2011})},\ \Eprint {http://arxiv.org/abs/1107.4642}
  {arXiv:1107.4642 [hep-th]} \BibitemShut {NoStop}%
\bibitem [{\citenamefont {{Grillo, C.}}\ \emph {et~al.}(2008)\citenamefont
  {{Grillo, C.}}, \citenamefont {{Lombardi, M.}},\ and\ \citenamefont {{Bertin,
  G.}}}]{Grillo:2008}%
  \BibitemOpen
  \bibfield  {author} {\bibinfo {author} {\bibnamefont {{Grillo, C.}}},
  \bibinfo {author} {\bibnamefont {{Lombardi, M.}}}, \ and\ \bibinfo {author}
  {\bibnamefont {{Bertin, G.}}},\ }\href {\doibase 10.1051/0004-6361:20077534}
  {\bibfield  {journal} {\bibinfo  {journal} {A\&A}\ }\textbf {\bibinfo
  {volume} {477}},\ \bibinfo {pages} {397} (\bibinfo {year}
  {2008})}\BibitemShut {NoStop}%
\bibitem [{\citenamefont {{Foreman-Mackey}}\ \emph {et~al.}(2013)\citenamefont
  {{Foreman-Mackey}}, \citenamefont {{Hogg}}, \citenamefont {{Lang}},\ and\
  \citenamefont {{Goodman}}}]{Emcee:2013}%
  \BibitemOpen
  \bibfield  {author} {\bibinfo {author} {\bibfnamefont {D.}~\bibnamefont
  {{Foreman-Mackey}}}, \bibinfo {author} {\bibfnamefont {D.~W.}\ \bibnamefont
  {{Hogg}}}, \bibinfo {author} {\bibfnamefont {D.}~\bibnamefont {{Lang}}}, \
  and\ \bibinfo {author} {\bibfnamefont {J.}~\bibnamefont {{Goodman}}},\ }\href
  {\doibase 10.1086/670067} {\bibfield  {journal} {\bibinfo  {journal} {pasp}\
  }\textbf {\bibinfo {volume} {125}},\ \bibinfo {pages} {306} (\bibinfo {year}
  {2013})},\ \Eprint {http://arxiv.org/abs/1202.3665} {arXiv:1202.3665
  [astro-ph.IM]} \BibitemShut {NoStop}%
\bibitem [{\citenamefont {Gelman}\ and\ \citenamefont
  {Rubin}(1992)}]{Gelman:1992}%
  \BibitemOpen
  \bibfield  {author} {\bibinfo {author} {\bibfnamefont {A.}~\bibnamefont
  {Gelman}}\ and\ \bibinfo {author} {\bibfnamefont {D.}~\bibnamefont {Rubin}},\
  }\href {\doibase 10.1103/PhysRevD.67.101301} {\bibfield  {journal} {\bibinfo
  {journal} {Statistical Science}\ }\textbf {\bibinfo {volume} {67}},\ \bibinfo
  {pages} {457} (\bibinfo {year} {1992})}\BibitemShut {NoStop}%
\bibitem [{\citenamefont {Akaike}(1974)}]{AIC:1974}%
  \BibitemOpen
  \bibfield  {author} {\bibinfo {author} {\bibfnamefont {H.}~\bibnamefont
  {Akaike}},\ }\href {\doibase 10.1109/TAC.1974.1100705} {\bibfield  {journal}
  {\bibinfo  {journal} {IEEE Transactions on Automatic Control}\ }\textbf
  {\bibinfo {volume} {19}},\ \bibinfo {pages} {716} (\bibinfo {year}
  {1974})}\BibitemShut {NoStop}%
\bibitem [{\citenamefont {Sugiura}(1978)}]{Sugiura:1978}%
  \BibitemOpen
  \bibfield  {author} {\bibinfo {author} {\bibfnamefont {N.}~\bibnamefont
  {Sugiura}},\ }\href {\doibase 10.1080/03610927808827599} {\bibfield
  {journal} {\bibinfo  {journal} {Communications in Statistics - Theory and
  Methods}\ }\textbf {\bibinfo {volume} {7}},\ \bibinfo {pages} {13} (\bibinfo
  {year} {1978})}\BibitemShut {NoStop}%
\bibitem [{\citenamefont {Schwarz}(1978)}]{schwarz1978}%
  \BibitemOpen
  \bibfield  {author} {\bibinfo {author} {\bibfnamefont {G.}~\bibnamefont
  {Schwarz}},\ }\href {\doibase 10.1214/aos/1176344136} {\bibfield  {journal}
  {\bibinfo  {journal} {Ann. Statist.}\ }\textbf {\bibinfo {volume} {6}},\
  \bibinfo {pages} {461} (\bibinfo {year} {1978})}\BibitemShut {NoStop}%
\bibitem [{\citenamefont {Demianski}\ \emph {et~al.}(2012)\citenamefont
  {Demianski}, \citenamefont {Piedipalumbo}, \citenamefont {Rubano},\ and\
  \citenamefont {Scudellaro}}]{Demianski:2012}%
  \BibitemOpen
  \bibfield  {author} {\bibinfo {author} {\bibfnamefont {M.}~\bibnamefont
  {Demianski}}, \bibinfo {author} {\bibfnamefont {E.}~\bibnamefont
  {Piedipalumbo}}, \bibinfo {author} {\bibfnamefont {C.}~\bibnamefont
  {Rubano}}, \ and\ \bibinfo {author} {\bibfnamefont {P.}~\bibnamefont
  {Scudellaro}},\ }\href {\doibase 10.1111/j.1365-2966.2012.21568.x} {\bibfield
   {journal} {\bibinfo  {journal} {Monthly Notices of the Royal Astronomical
  Society}\ }\textbf {\bibinfo {volume} {426}},\ \bibinfo {pages} {1396}
  (\bibinfo {year} {2012})}\BibitemShut {NoStop}%
\bibitem [{\citenamefont {Aviles}\ \emph {et~al.}(2012)\citenamefont {Aviles},
  \citenamefont {Gruber}, \citenamefont {Luongo},\ and\ \citenamefont
  {Quevedo}}]{Aviles:2012ay}%
  \BibitemOpen
  \bibfield  {author} {\bibinfo {author} {\bibfnamefont {A.}~\bibnamefont
  {Aviles}}, \bibinfo {author} {\bibfnamefont {C.}~\bibnamefont {Gruber}},
  \bibinfo {author} {\bibfnamefont {O.}~\bibnamefont {Luongo}}, \ and\ \bibinfo
  {author} {\bibfnamefont {H.}~\bibnamefont {Quevedo}},\ }\href {\doibase
  10.1103/PhysRevD.86.123516} {\bibfield  {journal} {\bibinfo  {journal} {Phys.
  Rev.}\ }\textbf {\bibinfo {volume} {D86}},\ \bibinfo {pages} {123516}
  (\bibinfo {year} {2012})},\ \Eprint {http://arxiv.org/abs/1204.2007}
  {arXiv:1204.2007 [astro-ph.CO]} \BibitemShut {NoStop}%
\bibitem [{\citenamefont {Aviles}\ \emph {et~al.}(2017)\citenamefont {Aviles},
  \citenamefont {Klapp},\ and\ \citenamefont {Luongo}}]{Aviles:2016wel}%
  \BibitemOpen
  \bibfield  {author} {\bibinfo {author} {\bibfnamefont {A.}~\bibnamefont
  {Aviles}}, \bibinfo {author} {\bibfnamefont {J.}~\bibnamefont {Klapp}}, \
  and\ \bibinfo {author} {\bibfnamefont {O.}~\bibnamefont {Luongo}},\ }\href
  {\doibase 10.1016/j.dark.2017.07.002} {\bibfield  {journal} {\bibinfo
  {journal} {Phys. Dark Univ.}\ }\textbf {\bibinfo {volume} {17}},\ \bibinfo
  {pages} {25} (\bibinfo {year} {2017})},\ \Eprint
  {http://arxiv.org/abs/1606.09195} {arXiv:1606.09195 [astro-ph.CO]}
  \BibitemShut {NoStop}%
\bibitem [{\citenamefont {Zhang}\ \emph {et~al.}(2017)\citenamefont {Zhang},
  \citenamefont {Li},\ and\ \citenamefont {Xia}}]{Zhang:2016urt}%
  \BibitemOpen
  \bibfield  {author} {\bibinfo {author} {\bibfnamefont {M.-J.}\ \bibnamefont
  {Zhang}}, \bibinfo {author} {\bibfnamefont {H.}~\bibnamefont {Li}}, \ and\
  \bibinfo {author} {\bibfnamefont {J.-Q.}\ \bibnamefont {Xia}},\ }\href
  {\doibase 10.1140/epjc/s10052-017-5005-4} {\bibfield  {journal} {\bibinfo
  {journal} {Eur. Phys. J.}\ }\textbf {\bibinfo {volume} {C77}},\ \bibinfo
  {pages} {434} (\bibinfo {year} {2017})},\ \Eprint
  {http://arxiv.org/abs/1601.01758} {arXiv:1601.01758 [astro-ph.CO]}
  \BibitemShut {NoStop}%
\bibitem [{\citenamefont {Nair}\ \emph {et~al.}(2012)\citenamefont {Nair},
  \citenamefont {Jhingan},\ and\ \citenamefont {Jain}}]{Nair_2012}%
  \BibitemOpen
  \bibfield  {author} {\bibinfo {author} {\bibfnamefont {R.}~\bibnamefont
  {Nair}}, \bibinfo {author} {\bibfnamefont {S.}~\bibnamefont {Jhingan}}, \
  and\ \bibinfo {author} {\bibfnamefont {D.}~\bibnamefont {Jain}},\ }\href
  {\doibase 10.1088/1475-7516/2012/01/018} {\bibfield  {journal} {\bibinfo
  {journal} {Journal of Cosmology and Astroparticle Physics}\ }\textbf
  {\bibinfo {volume} {2012}},\ \bibinfo {pages} {018} (\bibinfo {year}
  {2012})}\BibitemShut {NoStop}%
\bibitem [{\citenamefont {Al~Mamon}\ and\ \citenamefont
  {Das}(2016)}]{Sudipta:2016}%
  \BibitemOpen
  \bibfield  {author} {\bibinfo {author} {\bibfnamefont {A.}~\bibnamefont
  {Al~Mamon}}\ and\ \bibinfo {author} {\bibfnamefont {S.}~\bibnamefont {Das}},\
  }\href {\doibase 10.1142/S0218271816500322} {\bibfield  {journal} {\bibinfo
  {journal} {International Journal of Modern Physics D}\ }\textbf {\bibinfo
  {volume} {25}},\ \bibinfo {pages} {1650032} (\bibinfo {year} {2016})},\
  \Eprint {http://arxiv.org/abs/https://doi.org/10.1142/S0218271816500322}
  {https://doi.org/10.1142/S0218271816500322} \BibitemShut {NoStop}%
\bibitem [{\citenamefont {dos Santos}\ \emph {et~al.}(2016)\citenamefont {dos
  Santos}, \citenamefont {Reis},\ and\ \citenamefont {Waga}}]{Santos_2016}%
  \BibitemOpen
  \bibfield  {author} {\bibinfo {author} {\bibfnamefont {M.~V.}\ \bibnamefont
  {dos Santos}}, \bibinfo {author} {\bibfnamefont {R.}~\bibnamefont {Reis}}, \
  and\ \bibinfo {author} {\bibfnamefont {I.}~\bibnamefont {Waga}},\ }\href
  {\doibase 10.1088/1475-7516/2016/02/066} {\bibfield  {journal} {\bibinfo
  {journal} {Journal of Cosmology and Astroparticle Physics}\ }\textbf
  {\bibinfo {volume} {2016}},\ \bibinfo {pages} {066} (\bibinfo {year}
  {2016})}\BibitemShut {NoStop}%
\bibitem [{\citenamefont {Mohan}(2017)}]{Mohan:2017}%
  \BibitemOpen
  \bibfield  {author} {\bibinfo {author} {\bibfnamefont {S.~A. . M.~T.}\
  \bibnamefont {Mohan}, \bibfnamefont {N.D.J.}},\ }\href {\doibase
  10.1140/epjc/s10052-017-5428-y} {\bibfield  {journal} {\bibinfo  {journal}
  {Eur. Phys. J. C}\ }\textbf {\bibinfo {volume} {77}},\ \bibinfo {pages} {849}
  (\bibinfo {year} {2017})}\BibitemShut {NoStop}%
\bibitem [{\citenamefont {Mamon}(2018)}]{Abdulla-MPLA:2018}%
  \BibitemOpen
  \bibfield  {author} {\bibinfo {author} {\bibfnamefont {A.~A.}\ \bibnamefont
  {Mamon}},\ }\href {\doibase 10.1142/S0217732318500566} {\bibfield  {journal}
  {\bibinfo  {journal} {Modern Physics Letters A}\ }\textbf {\bibinfo {volume}
  {33}},\ \bibinfo {pages} {1850056} (\bibinfo {year} {2018})},\ \Eprint
  {http://arxiv.org/abs/https://doi.org/10.1142/S0217732318500566}
  {https://doi.org/10.1142/S0217732318500566} \BibitemShut {NoStop}%
\bibitem [{\citenamefont {Das}\ \emph {et~al.}(2018)\citenamefont {Das},
  \citenamefont {Mamon},\ and\ \citenamefont {Banerjee}}]{Das_2018}%
  \BibitemOpen
  \bibfield  {author} {\bibinfo {author} {\bibfnamefont {S.}~\bibnamefont
  {Das}}, \bibinfo {author} {\bibfnamefont {A.~A.}\ \bibnamefont {Mamon}}, \
  and\ \bibinfo {author} {\bibfnamefont {M.}~\bibnamefont {Banerjee}},\ }\href
  {\doibase 10.1088/1674-4527/18/11/131} {\bibfield  {journal} {\bibinfo
  {journal} {Research in Astronomy and Astrophysics}\ }\textbf {\bibinfo
  {volume} {18}},\ \bibinfo {pages} {131} (\bibinfo {year} {2018})}\BibitemShut
  {NoStop}%
\bibitem [{\citenamefont {Rom\'an-Garza}(2019)}]{Garza:2019}%
  \BibitemOpen
  \bibfield  {author} {\bibinfo {author} {\bibfnamefont {V.~T. M. J. e.~a.}\
  \bibnamefont {Rom\'an-Garza}, \bibfnamefont {J.}},\ }\href {\doibase
  10.1140/epjc/s10052-019-7390-3} {\bibfield  {journal} {\bibinfo  {journal}
  {Eur. Phys. J. C}\ }\textbf {\bibinfo {volume} {79}},\ \bibinfo {pages} {890}
  (\bibinfo {year} {2019})}\BibitemShut {NoStop}%
\bibitem [{\citenamefont {Hernández-Almada}\ \emph {et~al.}(2020)\citenamefont
  {Hernández-Almada}, \citenamefont {Leon}, \citenamefont {Magaña},
  \citenamefont {García-Aspeitia},\ and\ \citenamefont
  {Motta}}]{Almada-GEDE:2020}%
  \BibitemOpen
  \bibfield  {author} {\bibinfo {author} {\bibfnamefont {A.}~\bibnamefont
  {Hernández-Almada}}, \bibinfo {author} {\bibfnamefont {G.}~\bibnamefont
  {Leon}}, \bibinfo {author} {\bibfnamefont {J.}~\bibnamefont {Magaña}},
  \bibinfo {author} {\bibfnamefont {M.~A.}\ \bibnamefont {García-Aspeitia}}, \
  and\ \bibinfo {author} {\bibfnamefont {V.}~\bibnamefont {Motta}},\
  }\href@noop {} {\enquote {\bibinfo {title} {Generalized emergent dark energy:
  observational hubble data constraints and stability analysis},}\ } (\bibinfo
  {year} {2020}),\ \Eprint {http://arxiv.org/abs/2002.12881} {arXiv:2002.12881
  [astro-ph.CO]} \BibitemShut {NoStop}%
\bibitem [{\citenamefont {Li}\ and\ \citenamefont {Barrow}(2009)}]{Li:2009}%
  \BibitemOpen
  \bibfield  {author} {\bibinfo {author} {\bibfnamefont {B.}~\bibnamefont
  {Li}}\ and\ \bibinfo {author} {\bibfnamefont {J.~D.}\ \bibnamefont
  {Barrow}},\ }\href {\doibase 10.1103/PhysRevD.79.103521} {\bibfield
  {journal} {\bibinfo  {journal} {Phys. Rev. D}\ }\textbf {\bibinfo {volume}
  {79}},\ \bibinfo {pages} {103521} (\bibinfo {year} {2009})}\BibitemShut
  {NoStop}%
\bibitem [{\citenamefont {Velten}\ and\ \citenamefont
  {Schwarz}(2011)}]{Velten_2011}%
  \BibitemOpen
  \bibfield  {author} {\bibinfo {author} {\bibfnamefont {H.}~\bibnamefont
  {Velten}}\ and\ \bibinfo {author} {\bibfnamefont {D.~J.}\ \bibnamefont
  {Schwarz}},\ }\href {\doibase 10.1088/1475-7516/2011/09/016} {\bibfield
  {journal} {\bibinfo  {journal} {Journal of Cosmology and Astroparticle
  Physics}\ }\textbf {\bibinfo {volume} {2011}},\ \bibinfo {pages} {016}
  (\bibinfo {year} {2011})}\BibitemShut {NoStop}%
\bibitem [{\citenamefont {Velten}\ and\ \citenamefont
  {Schwarz}(2012)}]{Velten:2012uv}%
  \BibitemOpen
  \bibfield  {author} {\bibinfo {author} {\bibfnamefont {H.}~\bibnamefont
  {Velten}}\ and\ \bibinfo {author} {\bibfnamefont {D.}~\bibnamefont
  {Schwarz}},\ }\href {\doibase 10.1103/PhysRevD.86.083501} {\bibfield
  {journal} {\bibinfo  {journal} {Phys. Rev. D}\ }\textbf {\bibinfo {volume}
  {86}},\ \bibinfo {pages} {083501} (\bibinfo {year} {2012})},\ \Eprint
  {http://arxiv.org/abs/1206.0986} {arXiv:1206.0986 [astro-ph.CO]} \BibitemShut
  {NoStop}%
\bibitem [{\citenamefont {Velten}\ \emph {et~al.}(2014)\citenamefont {Velten},
  \citenamefont {Caramês}, \citenamefont {Fabris}, \citenamefont {Casarini},\
  and\ \citenamefont {Batista}}]{Velten:2014xca}%
  \BibitemOpen
  \bibfield  {author} {\bibinfo {author} {\bibfnamefont {H.}~\bibnamefont
  {Velten}}, \bibinfo {author} {\bibfnamefont {T.~R.~P.}\ \bibnamefont
  {Caramês}}, \bibinfo {author} {\bibfnamefont {J.~C.}\ \bibnamefont
  {Fabris}}, \bibinfo {author} {\bibfnamefont {L.}~\bibnamefont {Casarini}}, \
  and\ \bibinfo {author} {\bibfnamefont {R.~C.}\ \bibnamefont {Batista}},\
  }\href {\doibase 10.1103/PhysRevD.90.123526} {\bibfield  {journal} {\bibinfo
  {journal} {Phys. Rev. D}\ }\textbf {\bibinfo {volume} {90}},\ \bibinfo
  {pages} {123526} (\bibinfo {year} {2014})},\ \Eprint
  {http://arxiv.org/abs/1410.3066} {arXiv:1410.3066 [astro-ph.CO]} \BibitemShut
  {NoStop}%
\end{thebibliography}%


\end{document}